\documentclass{ws-ijtaf2}
\usepackage{subfigure}
\usepackage{xcolor}
\usepackage{hyperref}

\usepackage{placeins}
\usepackage[T1]{fontenc}   
\usepackage[utf8]{inputenc} 
\begin{document}

\markboth{Julia Kończal}{Pricing options on the cryptocurrency futures contracts}

\title{PRICING OPTIONS ON THE CRYPTOCURRENCY FUTURES CONTRACTS}

\author{JULIA KOŃCZAL}

\address{Faculty of Pure and Applied Mathematics, Hugo Steinhaus Center, \\Wrocław University of Science and Technology, \\
50-370 Wrocław, Poland\,\\
\email{julia.konczal@pwr.edu.pl} }

\maketitle

\begin{abstract}
The cryptocurrency options market is notable for its high volatility and lower liquidity compared to traditional markets. These characteristics introduce significant challenges to traditional option pricing methodologies. Addressing these complexities requires advanced models that can effectively capture the dynamics of the market. We explore which option pricing models are most effective in valuing cryptocurrency options. Specifically, we calibrate and evaluate the performance of the Black–-Scholes, Merton Jump Diffusion, Variance Gamma, Kou, Heston, and Bates models. Our analysis focuses on pricing vanilla options on futures contracts for Bitcoin (BTC) and Ether (ETH). We find that the Black--Scholes model exhibits the highest pricing errors. In contrast, the Kou and Bates models achieve the lowest errors, with the Kou model performing the best for the BTC options and the Bates model for ETH options. The results highlight the importance of incorporating jumps and stochastic volatility into pricing models to better reflect the behavior of these assets.
\end{abstract}

\keywords{Bitcoin; cryptocurrency; option pricing; volatility; jumps.}

\section{Introduction}

The introduction of cryptocurrencies has revolutionized the traditional approach to money and finance. The first cryptocurrency, Bitcoin, was introduced by an anonymous creator (or group of creators) under the pseudonym Satoshi Nakamoto \citep{nakamoto2008bitcoin}. Its goal was to create an alternative financial system based on decentralization. Since then, cryptocurrencies have gone from a niche idea to a global phenomenon. Subsequent projects, such as Ethereum, expanded their capabilities by introducing smart contracts and decentralized applications.

As the cryptocurrency market matured and its capitalization increased, there was a need for more advanced derivatives. Initially, simple buying and selling transactions on stock exchanges dominated, but products such as futures contracts were soon introduced. The Chicago Board Options Exchange (CBOE) and the Chicago Mercantile Exchange (CME) launched in 2017 the first regulated Bitcoin futures contracts. This marked a milestone in the recognition of cryptocurrencies as a serious investment vehicle. The next stage of development was the options on these contracts. This allowed investors to manage risk more effectively.
The development of derivatives such as futures and options has not only increased the liquidity of the cryptocurrency market, but also created new challenges for investors and analysts. Due to the high volatility of cryptocurrencies, and the unique technological features of blockchain, the valuation of such instruments has become much more complicated than in the case of traditional assets. 

There is extensive literature on the behavior of cryptocurrency prices. For example, \citet{kristoufek2015main} analyzed the potential drivers of Bitcoin prices, including fundamental, speculative, and technical factors, while specifically addressing the potential influence of the Chinese market.  \citet{chen2020bitcoin} predicted Bitcoin prices using machine learning and statistical methods, focusing on the impact of sample dimensions and data structures on model performance. \citet{chu2017garch} performed the first GARCH modeling of the seven most popular cryptocurrencies, including the evaluation of 12 GARCH models for each cryptocurrency. \citet{doi:10.1137/19M1263042} developed a term structure model with a zero short rate for cryptocurrency interest rates, derived price processes for crypto discount bonds and interest rate derivatives. 

Recent studies have also applied topological data analysis techniques to capture the complex structure and dynamics of cryptocurrency markets, highlighting their usefulness in modeling volatility and return predictability \citep{RUDKIN2023102759, rudkin2024bitcoin}. The differences between cryptocurrency and traditional financial markets were analyzed by studying volatility structures, cross-market correlations, and distributional and memory properties of Bitcoin, Ethereum, gold, major stock indices and other asset classes \citep{baur2018bitcoin, klein2018bitcoin, konczal2024tail, wkatorek2023cryptocurrencies}.

Crypto derivatives which allow traders to speculate on future price movements without owning the underlying assets are also an area of intensive research. \citet{madan2019advanced} investigated the dynamics of Bitcoin prices by analyzing vanilla options available on the market, calibrating a series of Markov models on the option surface. They evaluated the pricing performance and optimal risk-neutral parameters of these models, utilizing data sourced from unregulated exchanges. \citet{li2019bitcoin} proposed the integration of a multiple-input Long Short-Term Memory (LSTM)-based prediction model with the Black--Scholes model to improve Bitcoin option pricing by incorporating Blockchain statistics and social network trends.

The introduction of machine learning models, specifically regression-tree methods, for cryptocurrency option pricing to address unique market dynamics and inefficiencies was conducted by \citet{brini2024pricing}. Their study highlighted the superior adaptability of machine learning models to the complexities of cryptocurrency markets, emphasizing their potential to improve pricing accuracy in emerging asset classes. \citet{hou2020pricing} proposed a pricing mechanism for Bitcoin options based on stochastic volatility with a correlated jump model, comparing it to a flexible cojump model. Their findings highlighted the significant impact of price jumps and cojumps on options pricing, as confirmed by simulations and analyses of the implied volatility curve. Their study revealed that a notable portion of price jumps is anticorrelated with jumps in volatility, offering pioneering insights into the role of jumps in cryptocurrency markets and their importance in option pricing.  The relevance of modeling rare and extreme events is well known also for other financial instruments. 
In particular, catastrophe bonds (CAT bonds) are priced using catastrophe-loss models that focus on low-frequency, high-severity tail events, providing a natural analogy to jump-based approaches used for pricing cryptocurrency derivatives \citep{lane2000pricing}.

Our study examines a highly turbulent period for cryptocurrencies, which helps to better identify which models perform well under such conditions. Unlike many existing studies that primarily analyze short-term options traded on unregulated markets or rely on simulated data rather than actual market prices, our research focuses on long-term options from a regulated exchange. This distinction ensures that our findings are grounded in real market conditions, providing a more robust assessment of the model performance in cryptocurrency option pricing. We note that the model is calibrated using data from a single trading day, which limits the generalizability of the results to other market conditions. Furthermore, we price options on Bitcoin and Ether futures using a diverse set of models, including Black--Scholes, Merton Jump Diffusion, Heston, Kou, Variance Gamma, and Bates (Stochastic Volatility with Jumps). We calibrate each model separately for different maturities to better capture market dynamics across various time horizons. This allows for a more detailed evaluation of the effectiveness of the model.

This paper is structured as follows. In \autoref{sec:data} we explain the data under investigation, namely Bitcoin and Ether prices, as well as vanilla call options on Bitcoin and Ether futures contracts. In \autoref{sec:pricing} we calibrate six different pricing models: Black--Scholes, Variance Gamma, Merton Jump Diffusion, Kou, Heston, and Bates. The calibration is performed using market data for out-of-the-money options for eight maturities. We use calibrated models to separately price all available options for each maturity. \autoref{sec:errors} is devoted to the analysis of the performance of the models. We study the accuracy of the option pricing models by computing several error metrics: root mean squared error (RMSE), mean absolute error (MAE), mean absolute percentage error (MAPE), and mean squared logarithmic error (MSLE). These metrics are calculated separately for each maturity to assess the performance of the models across different time horizons. Additionally, we compute the error measures for all maturities combined to provide a summary of each model’s overall pricing accuracy. \autoref{sec:conclusions} concludes the paper.

\section{Data}
\label{sec:data}
In our study, we will focus on two major cryptocurrencies: Bitcoin (BTC) and Ether (ETH). Bitcoin is the first cryptocurrency, which was fully implemented and is currently the most recognized and valuable. The history of Bitcoin began, when a person under the pseudonym Satoshi Nakamoto published a white paper called ``Bitcoin: A Peer-to-Peer Electronic Cash System'' \citep{nakamoto2008bitcoin}.

Ethereum, by contrast, is a blockchain platform that enables the creation of decentralized applications and smart contracts \citep{wood2014ethereum}. The concept was first introduced by Vitalik Buterin at a conference in Miami \citep{buterin2014next}. Its native cryptocurrency, Ether, plays a key role within the Ethereum ecosystem \citep{jani2017overview}. It is the basic payment unit on the Ethereum platform, used to conduct transactions, pay for data processing on the blockchain, and pay for the execution of smart contracts. Technically, Ether is the cryptocurrency, while Ethereum refers to the underlying blockchain platform. However, in practice, the term ``Ethereum'' is often used interchangeably to describe both the network and the cryptocurrency itself.

\begin{figure}[h]
    \centering
    \subfigure[Bitcoin]{%
        \includegraphics[width=0.46\textwidth]{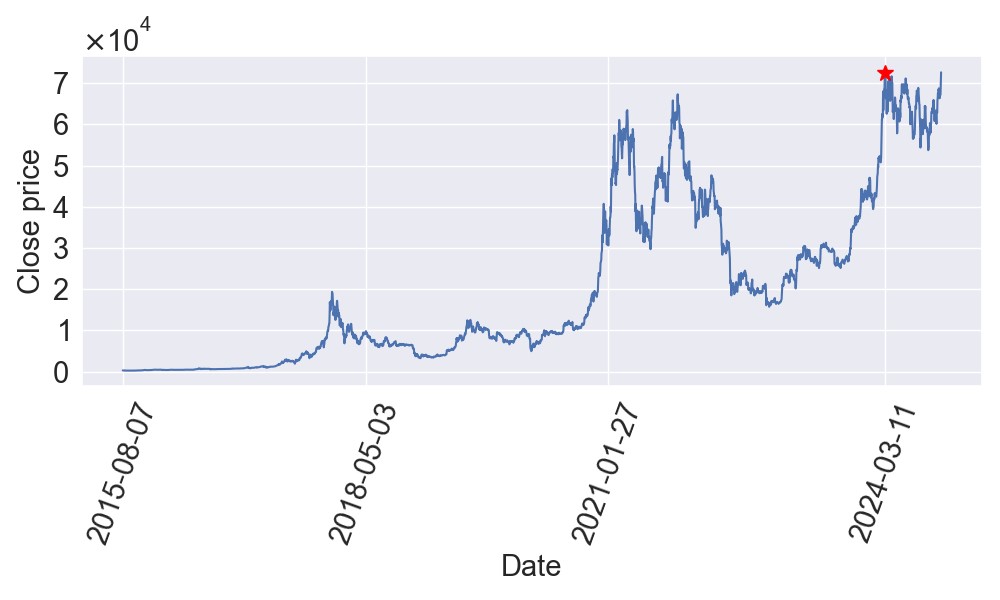}
        \label{fig:BTC_2025}
    }\hspace{0.05\textwidth}
    \subfigure[Ether]{%
        \includegraphics[width=0.46\textwidth]{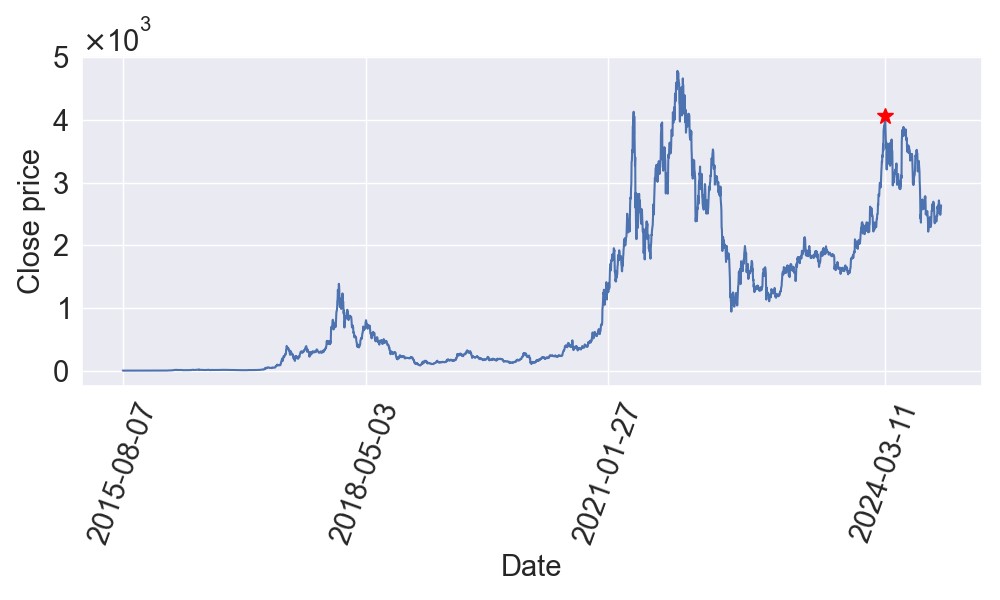}
        \label{fig:ETH_2025}
    }
    \caption{Closing prices of BTC and ETH for the period from 07.08.2015 to 29.10.2024. The red star marks the price on 11.03.2024, which corresponds to the trade date in our study.}
    \label{fig:prices}
\end{figure}

In \autoref{fig:prices} closing prices of both BTC and ETH are depicted. Both plots show similar shapes. The red star marks the price on 11.03.2024, which corresponds to the trade date in our study. On that day, the cryptocurrency market experienced significant events that impacted both Bitcoin and Ether. Bitcoin has reached a new price record, exceeding \$71,000. Ether also attracted attention as its price breached the \$4,000 mark for the first time since 2021.

To calibrate the models and assess pricing performance, we consider European call options written on Bitcoin and Ether futures contracts traded on a regulated derivatives exchange,  the Chicago Mercantile Exchange (CME).  The trade date is March 11, 2024, and we focus on expiration dates throughout 2024: June, July, August, September, December and 2025: March, June, December.  For BTC options, the number of available contracts used in the analysis for each maturity was 126, 125, 109, 113, 57, 53, 53, and 53, respectively. The strike prices started at 1000 USD and varied across maturities; for the June expiry they ranged up to 28000 USD, while for the December 2025 expiry the upper bound reached 32,000 USD. For ETH options, the corresponding numbers of contracts were 77, 76, 71, 69, 37, 37, 37, and 37, with strike prices beginning at 100 USD and extending up to 16000 USD for the June expiry and up to 18000 USD for the December 2025 expiry. The option contracts used in the analysis are standardized, each based on one futures contract of the respective underlying asset. Pricing units are denominated in U.S. dollars per Bitcoin or Ether, with minimum tick sizes defined by the exchange rules. Trading typically occurs five days a week, with brief daily breaks. Contract listings follow a regular cycle of near-term monthly expirations and longer-dated quarterly expirations. Trading terminates on the last Friday of the contract month, with adjustments if that day is not a business day in both London and the U.S.

For each option contract we use the official daily settlement prices published by the CME on the trade date. 
In our dataset, only the settlement price and the corresponding strike are available, intraday quotes, bid-ask quotes or transaction-level data are not accessible. 
Settlement prices can be viewed as proxies for unobservable mid-quotes, as they are determined by the exchange based on trading activity and order-book information near the close of the session. 
However, this also means that we cannot explicitly control for bid-ask spreads, transaction costs or temporary liquidity imbalances. 
As a consequence, part of the pricing errors reported in Section~\ref{sec:errors} may reflect microstructure noise and liquidity effects, in addition to pure model misspecification. 
This limitation should be kept in mind when interpreting the absolute magnitude of the errors.

It is important to note that the underlying assets in our analysis are futures contracts, not spot cryptocurrencies. Therefore, the valuation framework relies on the risk-neutral dynamics of the futures price, which already embeds the cost-of-carry relationship and market expectations. In contrast to traditional commodities, cryptocurrencies do not involve storage costs or convenience yield in the classical sense, but the futures–spot basis may still fluctuate due to funding rates, market segmentation, leverage conditions, and short-selling constraints. These factors can lead to persistent premiums or discounts of futures relative to spot prices. However, since the underlying of the options considered here is the futures contract itself, the basis does not directly affect the pricing formulas used in this study. It primarily influences the economic interpretation of market conditions on the valuation date.

\section{Pricing models and the results}
\label{sec:pricing}

Before proceeding with the pricing, we need to first calibrate our models. To this end, we calibrate each model individually for out-of-the-money (OTM) options at each maturity by minimizing the objective function, defined by the following formula \citep{oosterlee2019mathematical}:
\begin{equation}
\label{eq:calibration}
    \epsilon = \sum^{N_T}_{i=1} \sum^{N_K}_{j=1} \omega_{i,j} \left\{C(T_i,K_j) - C(\theta, T_i, K_j)\right\}^2,
\end{equation}
where $N_T$ -- the total number of maturities $T_i$ considered in the calibration process, $N_K$ -- the total number of strike prices $K_j$ considered for each maturity, $\omega_{i,j}$ -- the weighting factor applied to the difference between observed and model prices for maturity $T_i$ and strike $K_j$ (these weights can reflect the relative importance or reliability of specific data points like ATM option), $C(T_i, K_j)$ -- the observed market price of an option with maturity $T_i$ and strike $K_j$, $C(\theta, T_i, K_j)$ -- the model price of the same option, which depends on the model parameters $\theta$ being calibrated, and $\theta$ -- the vector of model parameters. 

 Although Equation
\eqref{eq:calibration} is written in a general form, in our empirical analysis the
calibration is performed separately for each maturity. This implies that
for every calibration run we set $N_T = 1$, so the objective function reduces to
a summation over the cross-section of strike prices for that specific maturity.
Consequently, the index $j$ refers to individual strikes within the chosen
maturity, while $i$ trivially identifies that maturity. We retain the general
notation for consistency with the standard formulation in the
literature. In our empirical implementation we assign equal weights to all options within a
given maturity, i.e. we set $\omega_{i,j} = 1$ for all strikes. This corresponds
to the standard unweighted least-squares calibration commonly used in the option
pricing literature.

For the calibration of all models we employ the Levenberg--Marquardt (LM)
algorithm, a standard numerical method for nonlinear least-squares problems \citep{marquardt1963algorithm}. The
LM algorithm can be interpreted as an interpolation between the Gauss--Newton
method and gradient descent: when the current iterate is far from the minimum,
the algorithm behaves similarly to gradient descent to ensure stability, whereas
in the vicinity of the optimum it approaches the Gauss--Newton update, allowing
for fast local convergence. This hybrid structure makes the LM procedure
particularly suitable for option-pricing calibration problems, where the
objective function is nonlinear but differentiable and may exhibit mild
non-convexity. In our implementation, we use the standard damping parameter
adaptation and iterate until changes in both the objective function and the
parameter vector fall below predefined tolerances.

In this study, we apply a closed-form solution for option pricing in the case of the Black--Scholes model and Merton Jump Diffusion model. For more complex models we derive option prices from the characteristic function \citep{oosterlee2019mathematical}. Each of the models describes the underlying asset price as a stochastic process. The price of the asset over time $S_t$ is described by the stochastic process $X_t$. The characteristic function $\varphi_X(u,t)$ of the stochastic process is defined as:
\begin{equation}
    \varphi_X(u,t) = \mathbb{E}[e^{iuX_t}].
\end{equation}
The characteristic function is specific to each of the models we study. To price an option using this method, we first define the stochastic process that describes the underlying asset, and then derive the characteristic function of that process. Once we have the characteristic function, we apply the Fourier transform to obtain the option price in the time domain.

In models for which no closed-form pricing formula exists, the Fourier inversion is performed using the \cite{carr1999option} Fast Fourier Transform (FFT) method. Following their approach, we compute the damped call transform
\[
\tilde{C}(k) = e^{\alpha k} C(k),
\]
with the damping factor fixed at $\alpha = 1.5$, which ensures integrability of the transform and numerical stability across all maturities and models considered.

The inversion is implemented on an equidistant frequency grid with $N = 2^{12} = 4096$ points. The spacing in the frequency domain is set to $\Delta u = 0.25$, which implies a log-strike spacing of 
\[
\Delta k = \frac{2\pi}{N \Delta u} \approx 0.0061.
\]
The integration range is truncated at $u_{\max} = (N-1)\Delta u$, a value we found sufficient for stable and accurate option prices. Increasing the grid size to $N = 8192$ or reducing the spacing to $\Delta u = 0.1$ produced changes below $10^{-6}$ in relative terms, confirming numerical robustness.
The Carr--Madan FFT method is applied consistently for all models relying on characteristic functions, including the Variance Gamma, Merton, Kou, Heston, and Bates models.

Throughout the paper all models are implemented under the risk-neutral measure. 
Since the underlying of each option is a futures contract, the risk-neutral drift of the futures price is zero, implying that the discounted futures price is a martingale. 
The parameters denoted by $\mu$ in the jump-diffusion and stochastic-volatility-with-jumps models refer to the risk-neutral parameters of the jump-size distribution (as in the original formulations of Merton, Kou, Heston and Bates), not to a physical drift. 
The compensator terms are included to ensure correct risk-neutral dynamics. In all models, we use the 3-month U.S. Treasury rate for discounting. Since the underlying is a futures contract, the interest rate affects only the discounting of option payoffs and does not enter the risk-neutral dynamics of the futures price.

\subsection{Black--Scholes model}

Black and Scholes \citeyearpar{black1973pricing} expressed the price process in their model as follows:
\begin{equation}
	 S_t = S_0 \exp \left\{\left(\mu -\frac{\sigma^2 }{2}\right)t + \sigma W_t \right\},
\end{equation}
\label{eq:gbm}where $S_t$ is the price of asset at time $t$, $\sigma$ is standard deviation of stock's returns $W_t$ is the Wiener process and $\sigma >0, \mu \in \mathbb{R}, S_0 >0$.  The value of the European call option in Black--Scholes model, which we utilise during options pricing, is expressed as follows:
\begin{equation}
\label{eq:bs}
	C_{BS}(S, K, T, t, r, \sigma) = N(d_+) S - N(d_-) K e^{-r(T - t)},
\end{equation}
where $N(x)$ -- standard normal cumulative distribution function, $K$ -- strike price of option, $r$ -- annualised risk-free interest rate, $ T$ -- time of option expiration, $d_{+}$ is denoted as:
\begin{equation}
	d_{+}={\frac {1}{\sigma {\sqrt {T-t}}}}\left\{\ln \left({\frac {S_{t}}{K}}\right)+\left(r+{\frac {\sigma ^{2}}{2}}\right)(T-t)\right\},
\end{equation}
and $	d_{-}$ is equivalent to:
\begin{equation}
	d_{-}=d_{+}-\sigma {\sqrt {T-t}}.
\end{equation}

\begin{figure}[h]
    \centering
    \subfigure[Bitcoin]{%
        \includegraphics[width=0.46\textwidth]{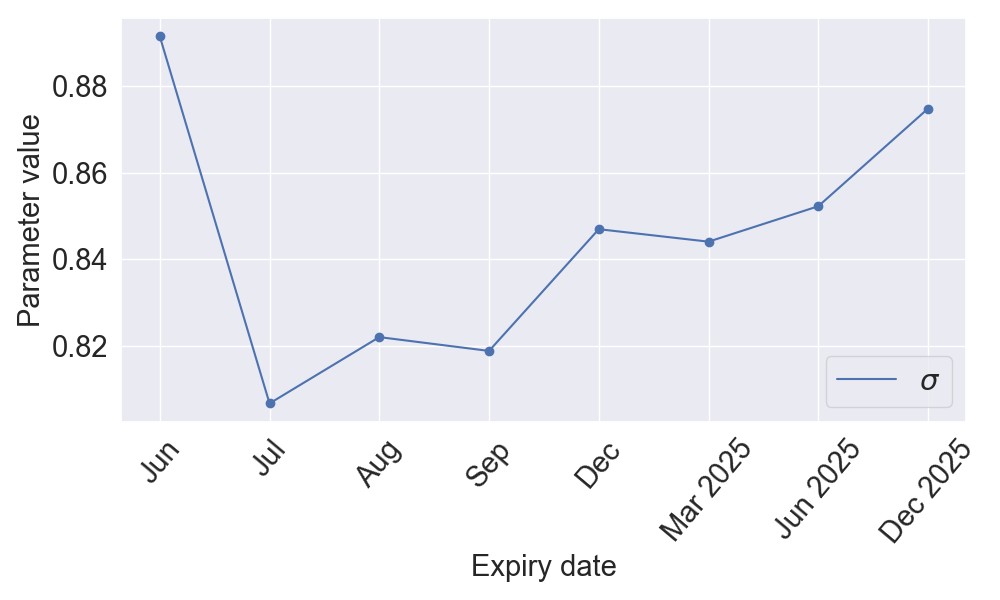}
        \label{fig:BS_param_BTC}
    }\hspace{0.05\textwidth}
    \subfigure[Ether]{%
        \includegraphics[width=0.46\textwidth]{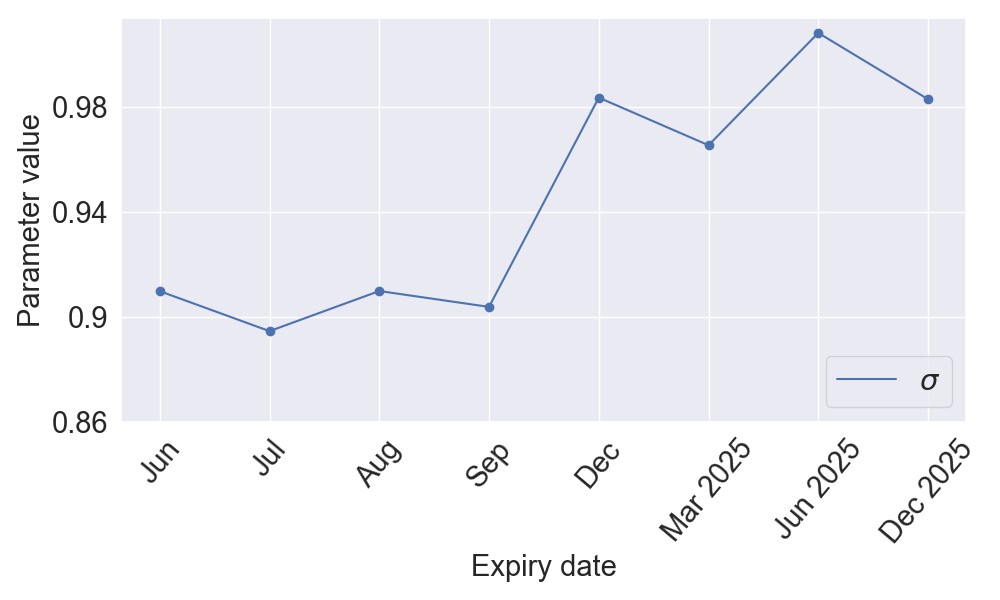}
        \label{fig:BS_param_ETH}
    }
    \caption{The optimal parameter values for each maturity calibrated using the Black--Scholes model.}
    \label{fig:BS_param}
\end{figure}

In \autoref{fig:BS_param} we present the results of the calibration for all the expiration dates of BTC and ETH. As we can observe, for BTC the values oscillate between $0.8$ and $0.9$, and for ETH the values are increasing starting from $0.91$ and exceeding $1$. These values suggest that the cryptocurrency options market is extremely volatile, but also they indicate large price fluctuations, which is consistent with historical observations for cryptocurrencies. We notice that the results are very high compared to the traditional assets. Assuming that the implied volatility for S\&P 500 options typically lies in the range of 0.15 to 0.20, the estimated volatilities for cryptocurrency options are approximately four to six times higher.
This observation aligns with empirical findings. In \cite{nzokem2024bitcoin} authors report that Bitcoin exhibits volatility levels approximately five times higher than those of the S\&P 500 in the short term and around four times higher in the long term.

	\begin{figure}[h]
    \centering
    \subfigure[June 2024]{%
        \includegraphics[width=0.29\textwidth]{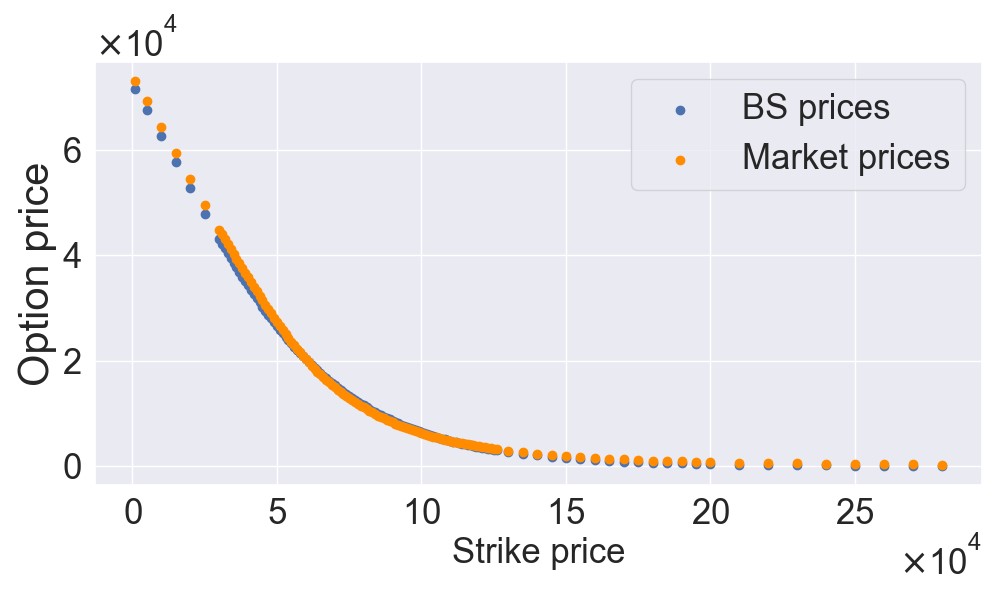}
        \label{fig:BS_Jun_BTC}
    }\hspace{0.03\textwidth}
    \subfigure[December 2024]{%
        \includegraphics[width=0.29\textwidth]{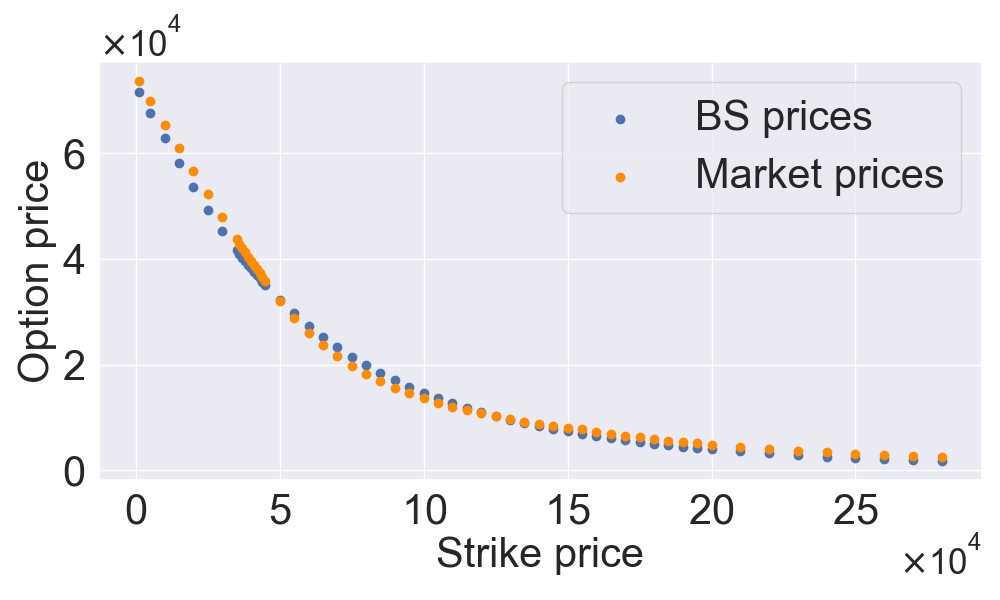}
        \label{fig:BS_Dec_BTC}
    }\hspace{0.03\textwidth}
    \subfigure[December 2025]{%
        \includegraphics[width=0.29\textwidth]{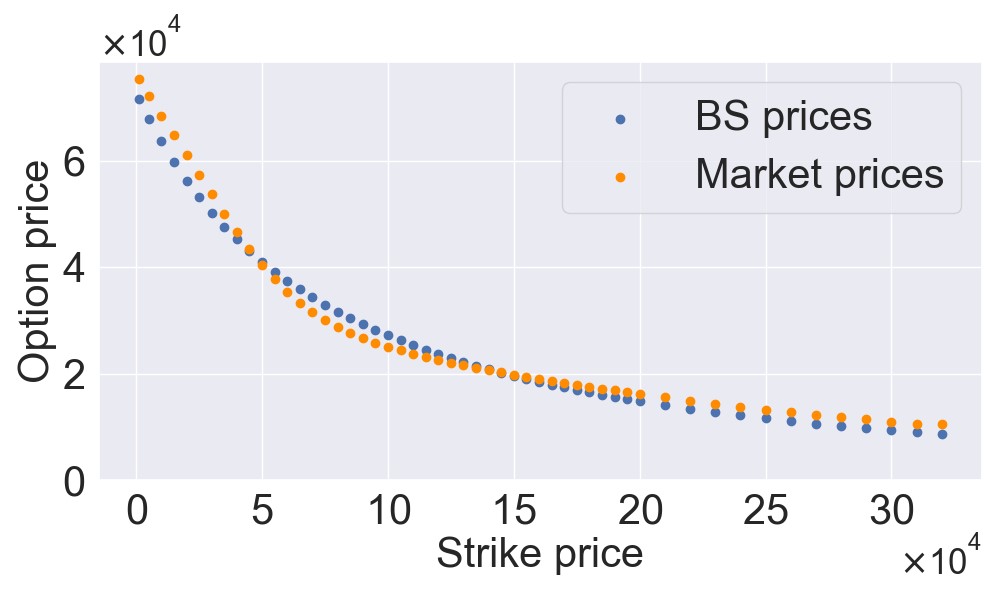}
        \label{fig:BS_Dec_2025_BTC}
    }
    \caption{Pricing results using the Black--Scholes model for options on BTC futures contracts with expiry dates in June 2024, December 2024, and December 2025.}
    \label{fig:BS_BTC}
\end{figure}

\begin{figure}[h]
    \centering
    \subfigure[June 2024]{%
        \includegraphics[width=0.29\textwidth]{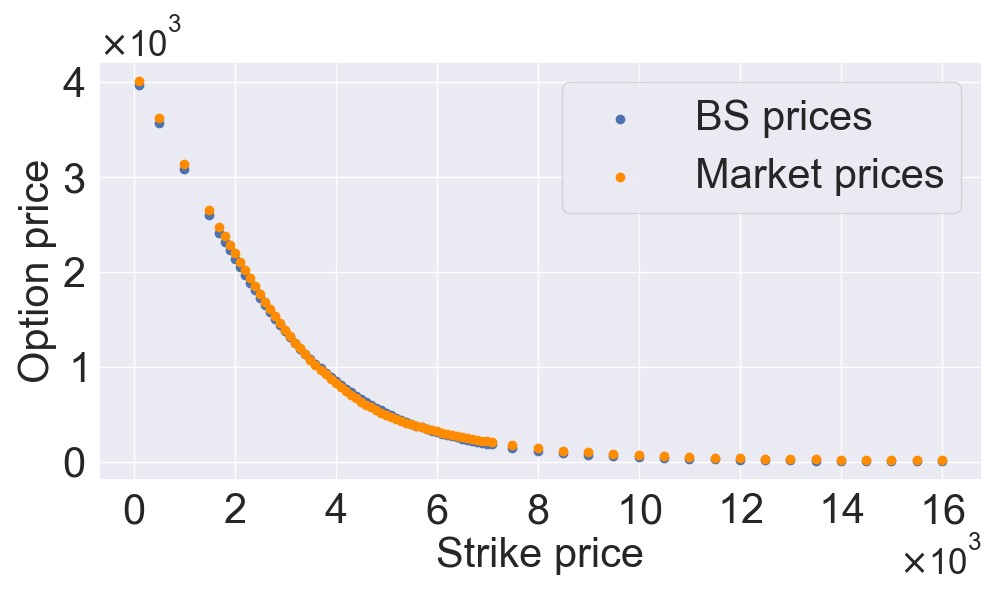}
        \label{fig:BS_Jun_ETH}
    }\hspace{0.03\textwidth}
    \subfigure[December 2024]{%
        \includegraphics[width=0.29\textwidth]{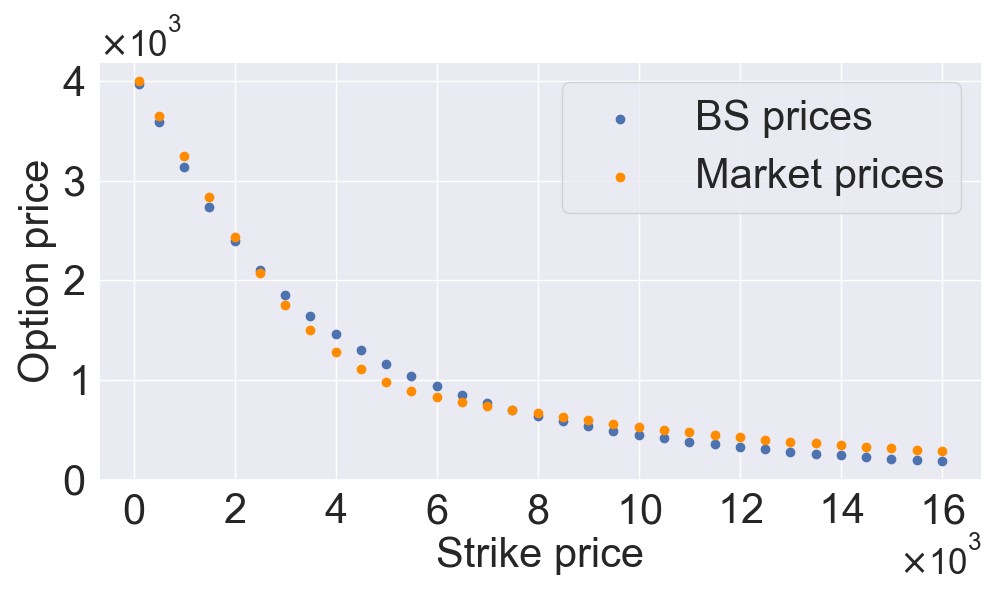}
        \label{fig:BS_Dec_ETH}
    }\hspace{0.03\textwidth}
    \subfigure[December 2025]{%
        \includegraphics[width=0.29\textwidth]{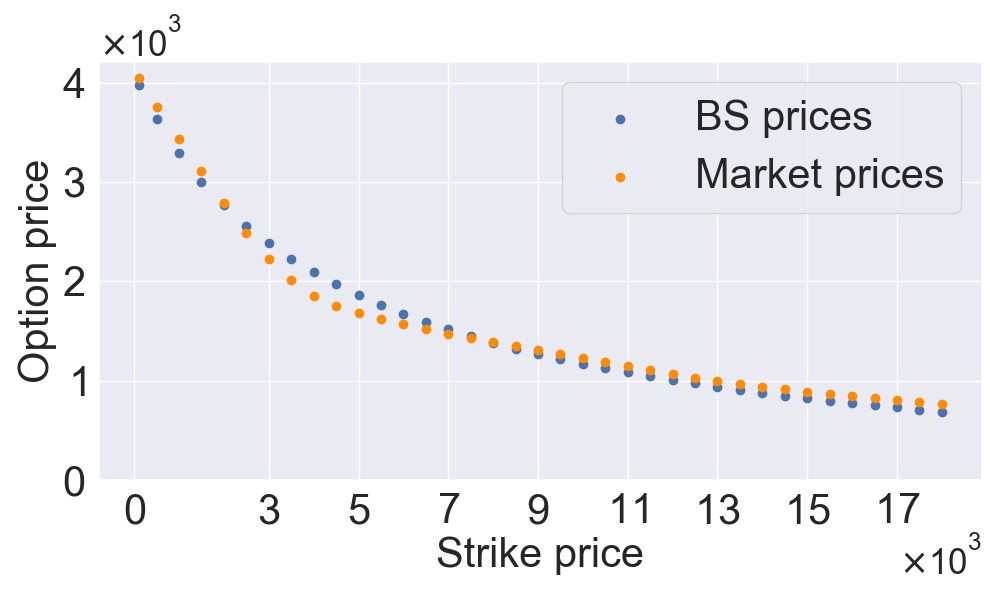}
        \label{fig:BS_Dec_2025_ETH}
    }
    \caption{Pricing results using the Black--Scholes model for options on ETH futures contracts with expiry dates in June 2024, December 2024, and December 2025.}
    \label{fig:BS_ETH}
\end{figure}

    In \autoref{fig:BS_BTC} and \autoref{fig:BS_ETH} we present pricing results for three expiration dates compared to the market data,  for BTC and ETH respectively. A visual examination suggests that the BS model does not perform well for longer maturities in both cases. The BS model is one of the most commonly used models for pricing options, but its basic assumptions often make it imprecise enough for assets characterized by high volatility.  When comparing the statistical differences and tail behavior we notice that cryptocurrencies differ from traditional commodities \citep{konczal2024tail}. Cryptocurrencies may be susceptible to more extreme price events, which requires the use of more complex stochastic models.

\subsection{Variance Gamma model}

\begin{figure}[h]
    \centering
    \subfigure[Bitcoin]{%
        \includegraphics[width=0.46\textwidth]{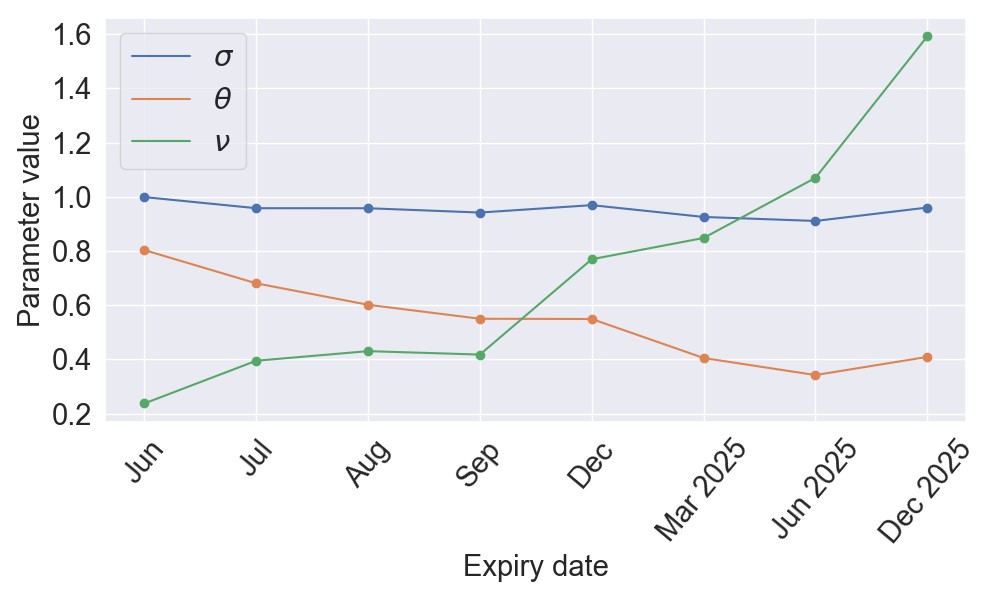}
        \label{fig:VG_param_BTC}
    }\hspace{0.05\textwidth}
    \subfigure[Ether]{%
        \includegraphics[width=0.46\textwidth]{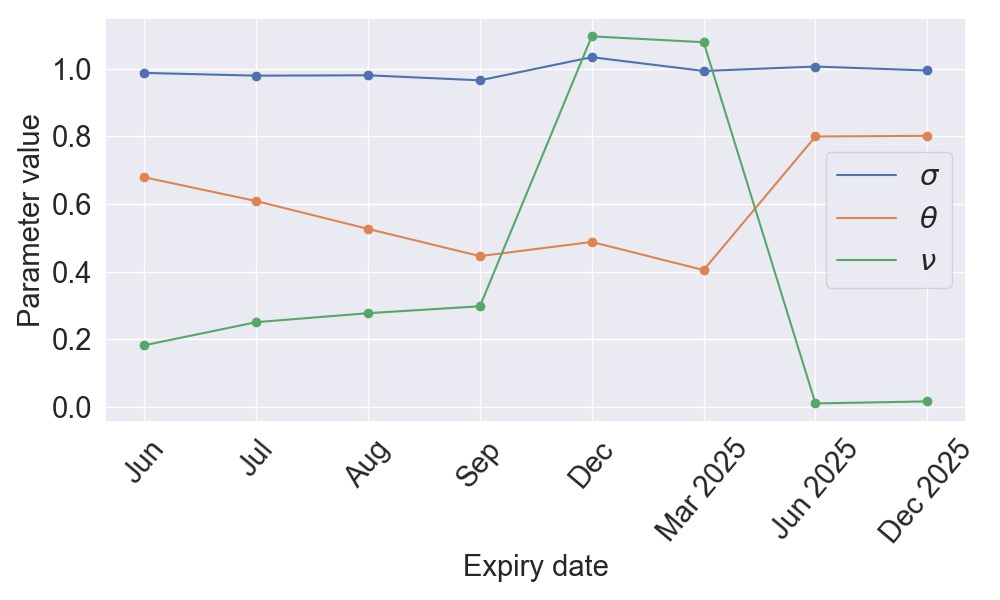}
        \label{fig:VG_param_ETH}
    }
    \caption{The optimal parameter values for each maturity calibrated using the Variance Gamma model.}
    \label{fig:VG_param}
\end{figure}

To better capture the characteristics of the cryptocurrency market, we now apply the Variance Gamma (VG) model, which allows for a more flexible modeling of price dynamics \citep{madan1998variance}. The VG model allows for simulating asymmetry in the distribution of returns and a better adjustment to market volatility, which makes it a more adequate tool for pricing cryptocurrency options. In this model the price process is expressed as follows:
\begin{equation}
    S_t = S_0    \frac{\exp(X_t)}{\mathbb{E}[\exp(X_t)]},
\end{equation}
where $S_t$ is the price of asset at time $t$, and  $X_t$ is a variance gamma process. We note that variance gamma process can be written as the difference of two independent gamma processes:
\begin{equation}
     X^{VG}(t;\sigma ,\nu ,\theta )\;=\;G (t;\mu _{p},\mu _{p}^{2}\,\nu )-G(t;\mu _{q},\mu _{q}^{2}\,\nu ),
\end{equation}
where
\begin{equation}
    \begin{cases}
    \mu_p = \frac{1}{2} \sqrt{\theta^2 + \frac{2\sigma^2}{\nu}} + \frac{\theta}{2} \quad \quad \\
    \mu_q = \frac{1}{2} \sqrt{\theta^2 + \frac{2\sigma^2}{\nu}} - \frac{\theta}{2}.
\end{cases}
\end{equation}
Here, $G(t; \mu, \lambda)$ is the gamma process -- a pure-jump, increasing Lévy process with independent and stationary increments. Its marginal distribution at time $t$ is a gamma distribution with mean $\mu t / \lambda$ and variance $\mu^2 t / \lambda$, where $\mu$ is the scale parameter and $\lambda$ controls the rate of jump arrivals.
The characteristic function for this model is expressed as:
\begin{equation}
\varphi_X(u, t) = \left( 1 - i \theta \nu u + \frac{1}{2} \sigma^2 \nu u^2 \right)^{-t/\nu}.
\end{equation}
Hence, we calibrate three parameters: $\sigma$ -- volatility, $\theta$ -- skewness, and $\nu$ -- kurtosis. The results of the calibration are presented in \autoref{fig:VG_param}. For BTC we observe high values of volatility, which are close to 1 for all of the maturities. The values of $\theta$ range between $0.35$ and $0.8$,  suggesting a noticeable skewness in the return distribution. We also see that $\nu$ is strictly increasing, starting from $0.22$ and exceeding $1.6$. A higher value of $\nu$ suggests leptokurtosis, which is a characteristic often associated with assets experiencing sudden price fluctuations. For ETH we observe analogous calibration results. The green curves, which represent the $\nu$ parameter, show a clear difference between the two assets. For BTC, $\nu$ increases gradually over time, which means that the chance of very large price moves increases for longer time horizons. In contrast, ETH shows a rising $\nu$ until mid-horizon, followed by a sharp drop, suggesting that the risk of extreme price changes is expected to decrease in the long term.

\begin{figure}[h]
    \centering
    \subfigure[June 2024]{%
        \includegraphics[width=0.29\textwidth]{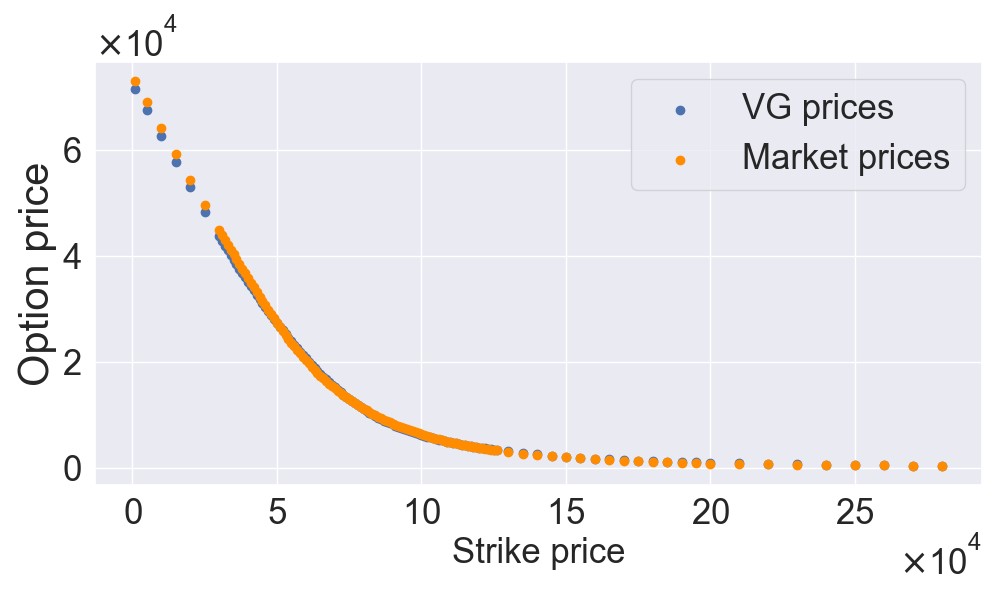}
        \label{fig:VG_Jun_BTC}
    }\hspace{0.03\textwidth}
    \subfigure[December 2024]{%
        \includegraphics[width=0.29\textwidth]{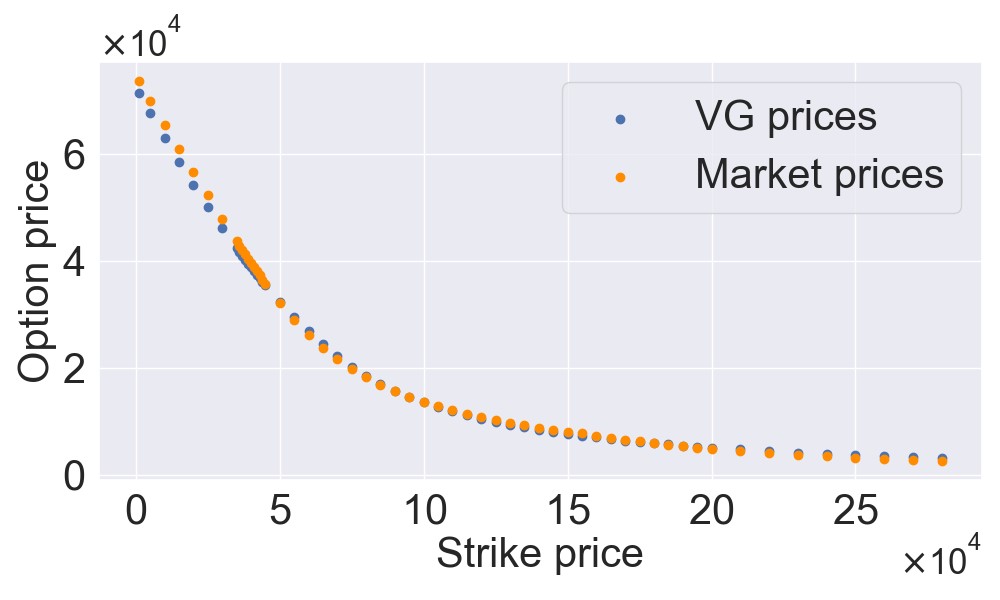}
        \label{fig:VG_Dec_BTC}
    }\hspace{0.03\textwidth}
    \subfigure[December 2025]{%
        \includegraphics[width=0.29\textwidth]{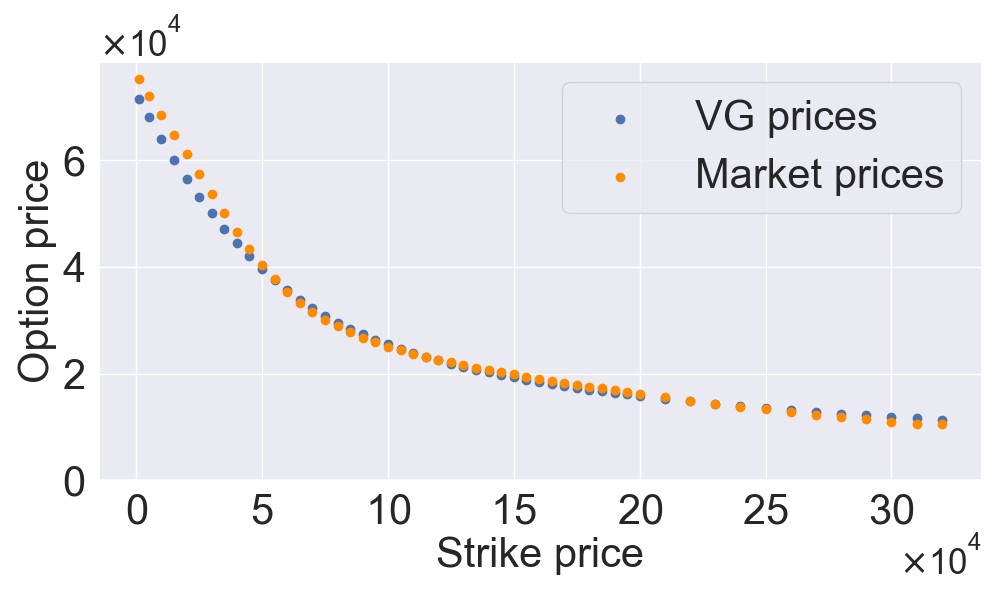}
        \label{fig:VG_Dec_2025_BTC}
    }
    \caption{Pricing results using the Variance Gamma model for options on BTC futures contracts with expiry dates in June 2024, December 2024, and December 2025.}
    \label{fig:VG_BTC}
\end{figure}

\begin{figure}[h]
    \centering
    \subfigure[June 2024]{%
        \includegraphics[width=0.29\textwidth]{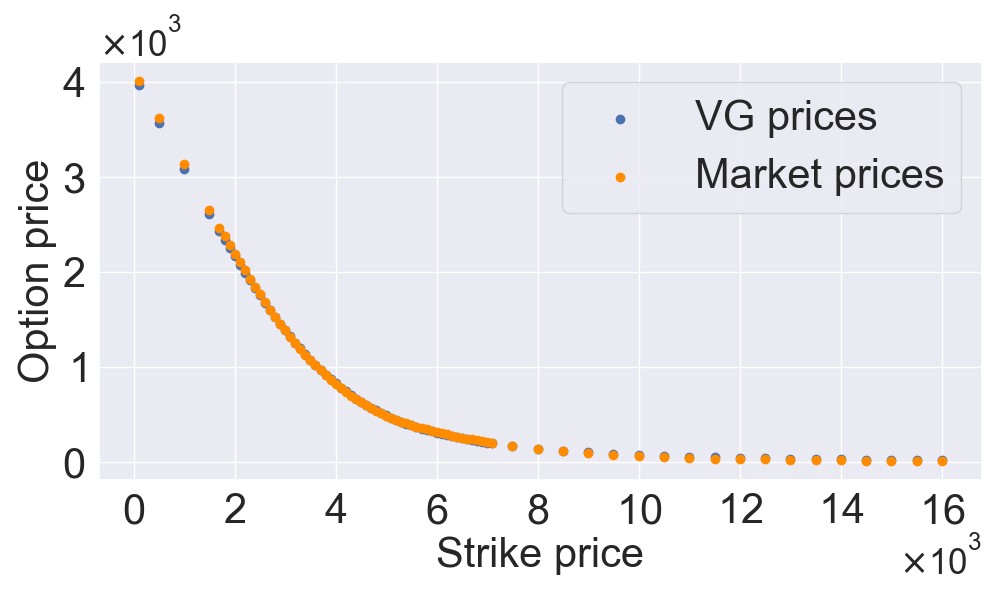}
        \label{fig:VG_Jun_ETH}
    }\hspace{0.03\textwidth}
    \subfigure[December 2024]{%
        \includegraphics[width=0.29\textwidth]{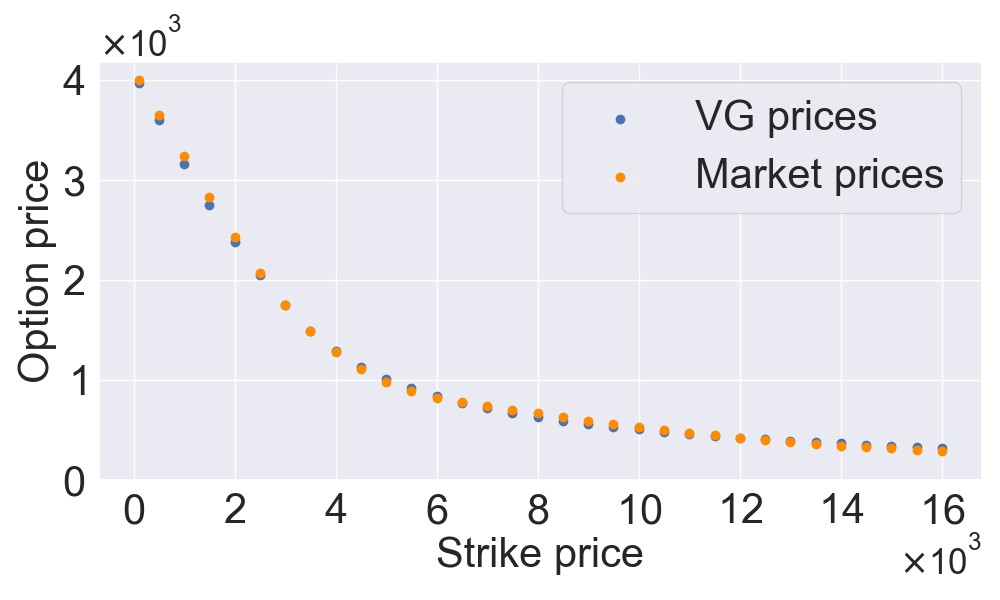}
        \label{fig:VG_Dec_ETH}
    }\hspace{0.03\textwidth}
    \subfigure[December 2025]{%
        \includegraphics[width=0.29\textwidth]{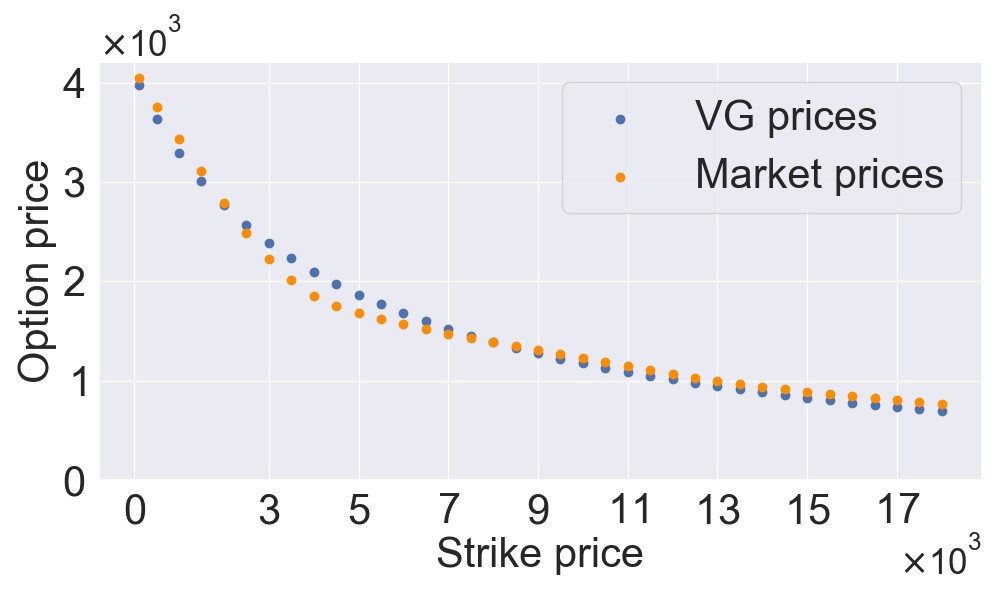}
        \label{fig:VG_Dec_2025_ETH}
    }
    \caption{Pricing results using the Variance Gamma model for options on ETH futures contracts with expiry dates in June 2024, December 2024, and December 2025.}
    \label{fig:VG_ETH}
\end{figure}

In \autoref{fig:VG_BTC} and \autoref{fig:VG_ETH}, we present the pricing results for three expiration dates, comparing them against market data for BTC and ETH. It is evident that for longer-dated options, the VG model outperforms the BS model.
However, for options maturing in December 2025 we can clearly observe differences between the model prices and market prices. This suggests that while the VG model provides a more accurate representation of option prices over extended horizons, further extensions may still be necessary to fully capture the market dynamics across all strike levels.

\subsection{Merton Jump Diffusion model}

\begin{figure}[h]
    \centering
    \subfigure[Bitcoin]{%
        \includegraphics[width=0.45\textwidth]{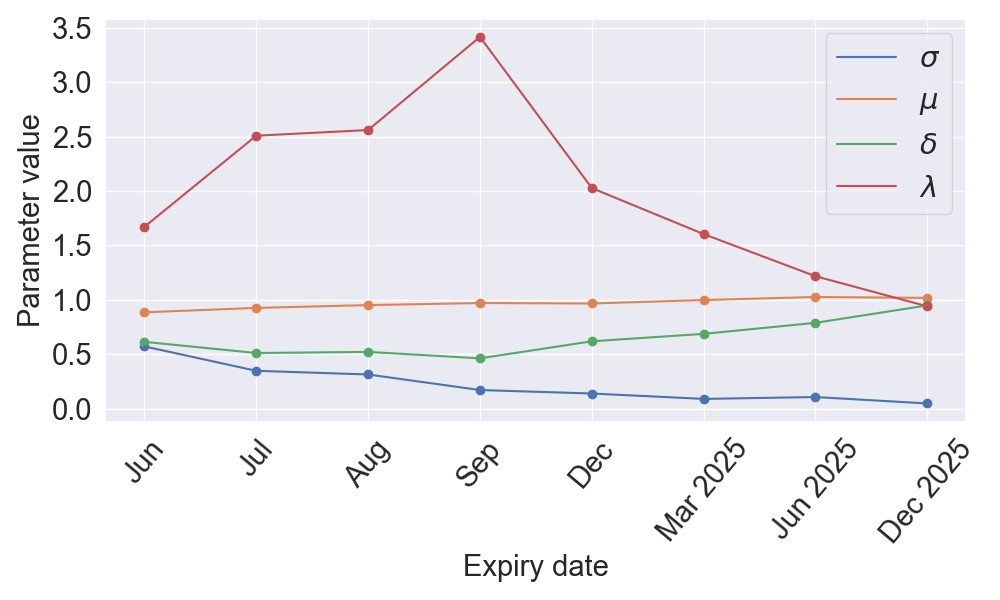}
        \label{fig:Merton_param_BTC}
    }\hspace{0.05\textwidth}
    \subfigure[Ether]{%
        \includegraphics[width=0.45\textwidth]{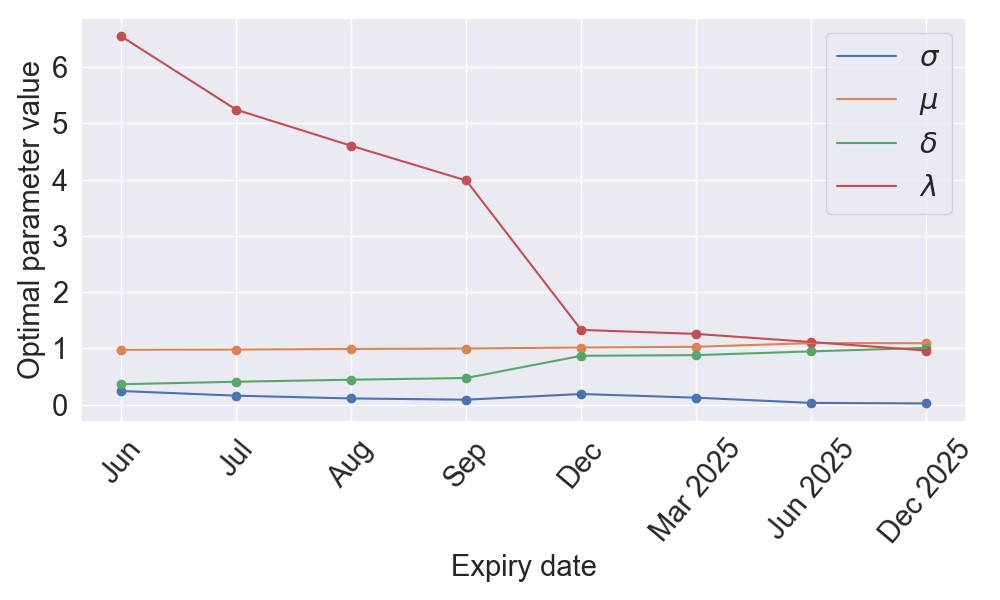}
        \label{fig:Merton_param_ETH}
    }
    \caption{The optimal parameter values for each maturity calibrated using the Merton Jump Diffusion model.}
    \label{fig:MJD_param}
\end{figure}

Next, we take a look at Merton Jump Diffusion (MJD) model \citep{merton1976option}. This is an extension of the classic BS model, introducing an additional jump component to the standard diffusion process. The price process is expressed as follows:
\begin{equation}
    S_t = S_0 \exp \left\{ \left(\alpha - \frac{\sigma^2}{2} - \lambda k \right) t + \sigma W_t + \sum_{i=1}^{N_t} Y_i \right\},
\end{equation}
where $S_t$ -- price of asset at time $t$, $W_t$ -- a Wiener process, $\sum_{i=1}^{N_t} Y_i$ -- a
compound Poisson jump process, $N_t$ -- Poisson process with intensity $\lambda$, $\alpha$ -- expected return on the asset, $\sigma$ --
volatility, $\ln(y_t) = Y_t$, and 
\begin{equation}
    \begin{cases}
        \ln(y_t) \sim \textit{i.i.d.} \ \mathcal{N}(\mu, \delta^2), \\
        \mathbb{E}[y_t - 1] = e^{\mu + \frac{1}{2} \delta^2} - 1 = k, \\
        \mathbb{E}\left[(y_t - 1 - \mathbb{E}[y_t - 1])^2 \right] = e^{2\mu + \delta^2} \left(e^{\delta^2} - 1\right).
    \end{cases} 
\end{equation}
The closed form solution of call option price can be derived as follows:
\begin{equation}
	C_{MJD}(S, K, T, t, r, \sigma, \mu, v, \lambda) = \sum_{k=0}^{\infty} \frac{e^{-\lambda \mu (T - t)} (\lambda \mu (T - t))^k}{k!} \, C_{BS}(S, K, T, t, r_k, \sigma_k),
\end{equation}
where 
\begin{equation}
	\sigma_k = \sqrt{\sigma^2 + k \tfrac{v^2}{T}},
\end{equation}
\begin{equation}
	r_k = r - \lambda(\mu-1) + \tfrac{k \ln{(\mu)}}{T},
\end{equation}
and $C_{BS} (S, K, T, t, r_k, \sigma_k)$ is the closed form solution of call option in Black--Scholes model, which was presented in Equation \eqref{eq:bs}.
Hence, we perform calibration of four parameters: $\sigma$, $\lambda$, $\mu$, $\delta$. The results are presented in  \autoref{fig:MJD_param}. For BTC, we observe that 
$\sigma$ decreases over time, starting from 0.7 and reaching 0.05 for options maturing in December 2025. The values of parameter
$\mu$ remain stable, oscillating around 0.95 across all expiration dates, suggesting that the typical magnitude of price jumps is consistently high, regardless of maturity.
The jump standard deviation 
$\delta$ shows an increasing trend, starting at approximately 0.4 and exceeding 0.94. It indicates growing uncertainty in jump sizes over longer time horizons. The jump intensity parameter 
$\lambda$ fluctuates between 1 and 3.5, suggesting that while jumps play a key role in BTC price dynamics, their occurrence rate varies across different maturities. For ETH, we observe very similar results. However, we notice significantly larger range of the jump intensity parameter 
$\lambda$, which varies between 1 and 7. This suggests that jumps occur more frequently in ETH compared to BTC.

\begin{figure}[h]
    \centering
    \subfigure[June 2024]{%
        \includegraphics[width=0.29\textwidth]{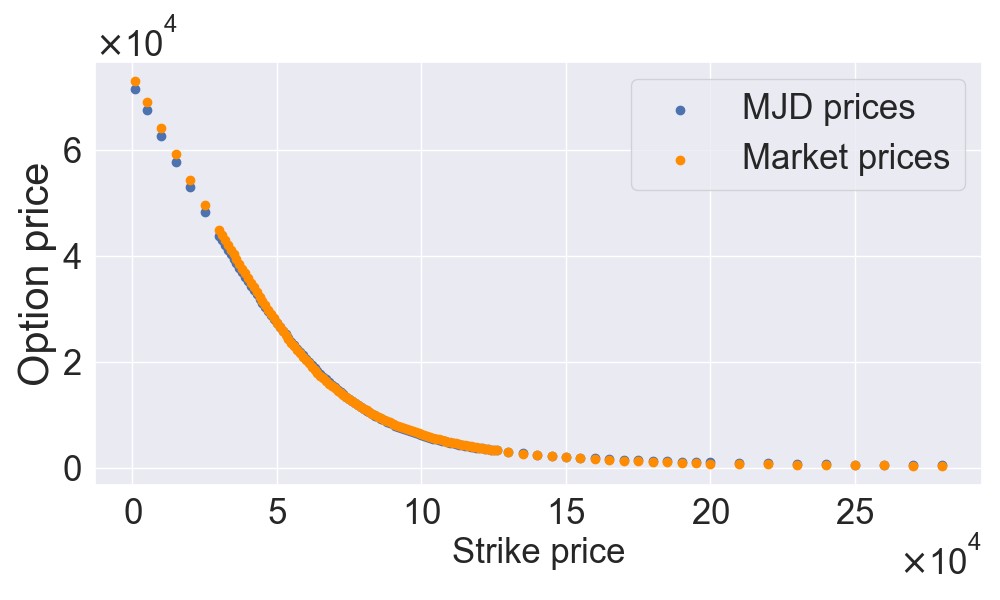}
        \label{fig:Merton_Jun_BTC}
    }\hspace{0.03\textwidth}
    \subfigure[December 2024]{%
        \includegraphics[width=0.29\textwidth]{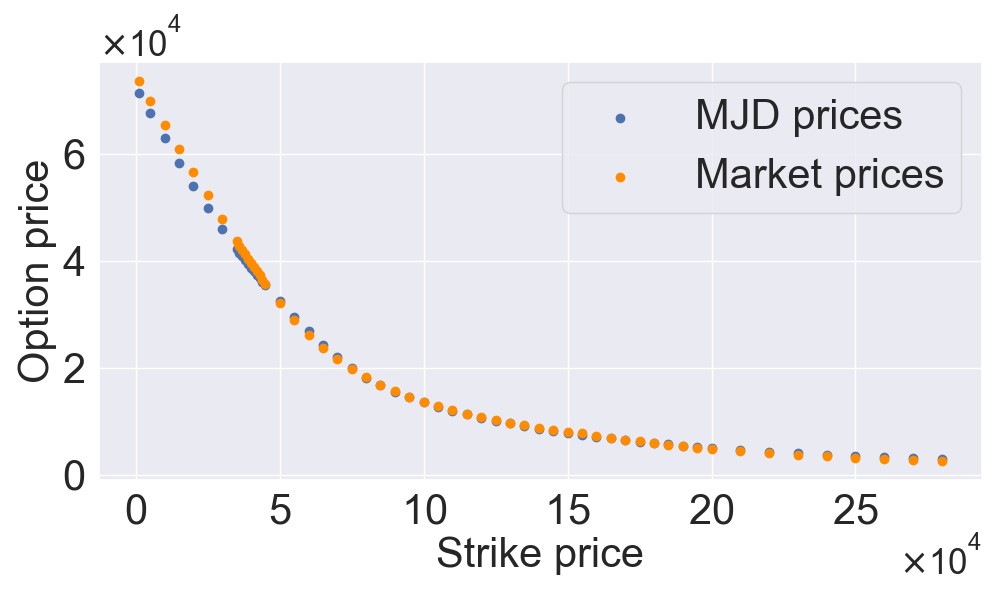}
        \label{fig:Merton_Dec_BTC}
    }\hspace{0.03\textwidth}
    \subfigure[December 2025]{%
        \includegraphics[width=0.29\textwidth]{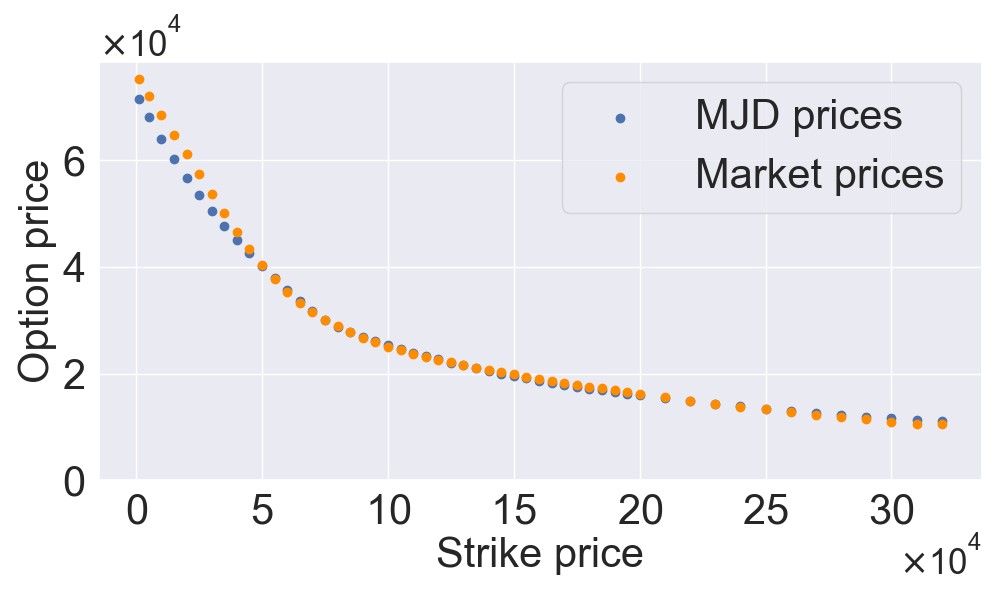}
        \label{fig:Merton_Dec_2025_BTC}
    }
    \caption{Pricing results using the Merton Jump Diffusion model for options on BTC futures contracts with expiry dates in June 2024, December 2024, and December 2025.}
    \label{fig:Merton_BTC}
\end{figure}

\begin{figure}[h]
    \centering
    \subfigure[June 2024]{%
        \includegraphics[width=0.29\textwidth]{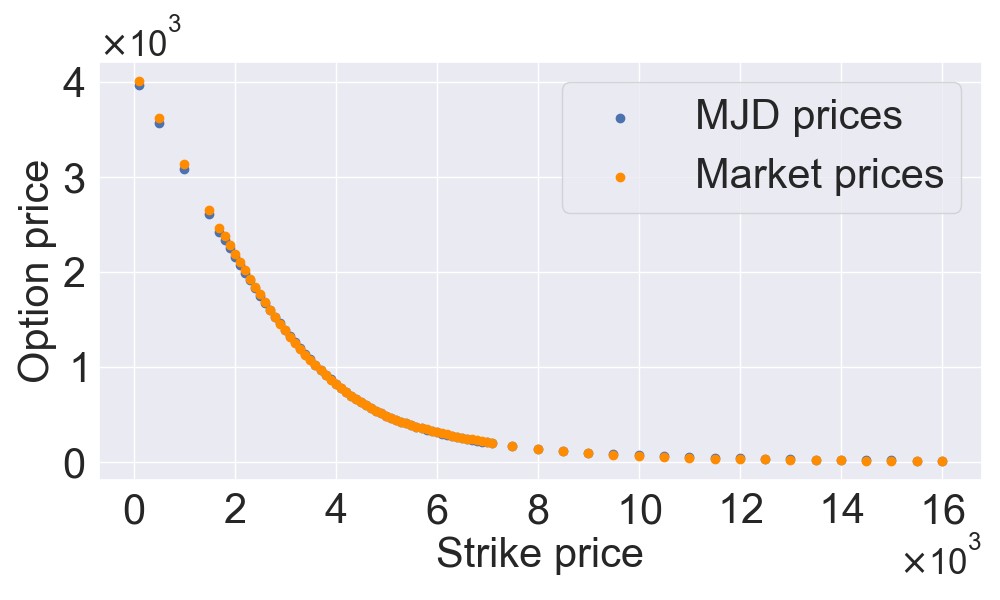}
        \label{fig:Merton_Jun_ETH}
    }\hspace{0.03\textwidth}
    \subfigure[December 2024]{%
        \includegraphics[width=0.29\textwidth]{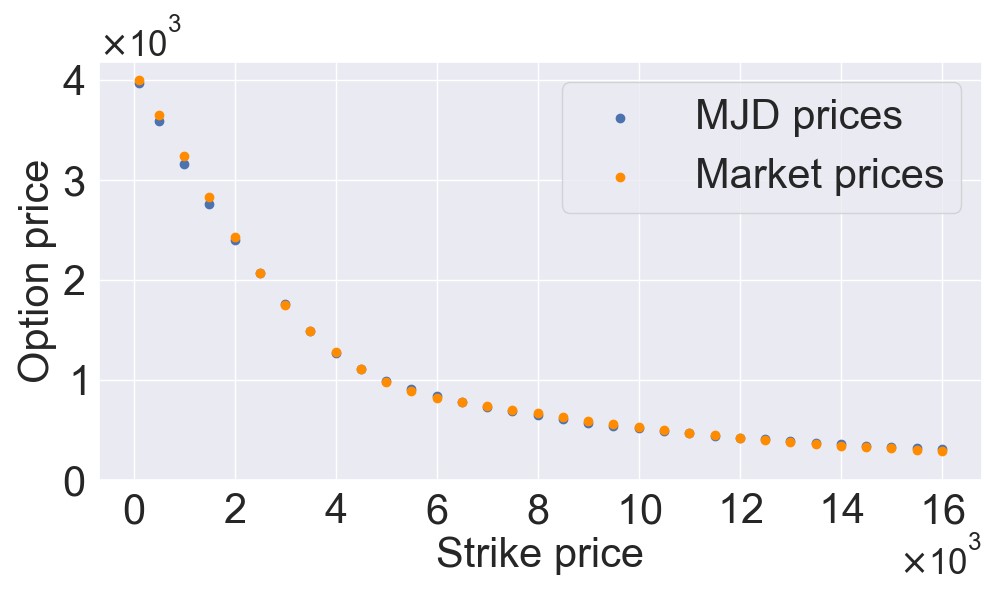}
        \label{fig:Merton_Dec_ETH}
    }\hspace{0.03\textwidth}
    \subfigure[December 2025]{%
        \includegraphics[width=0.29\textwidth]{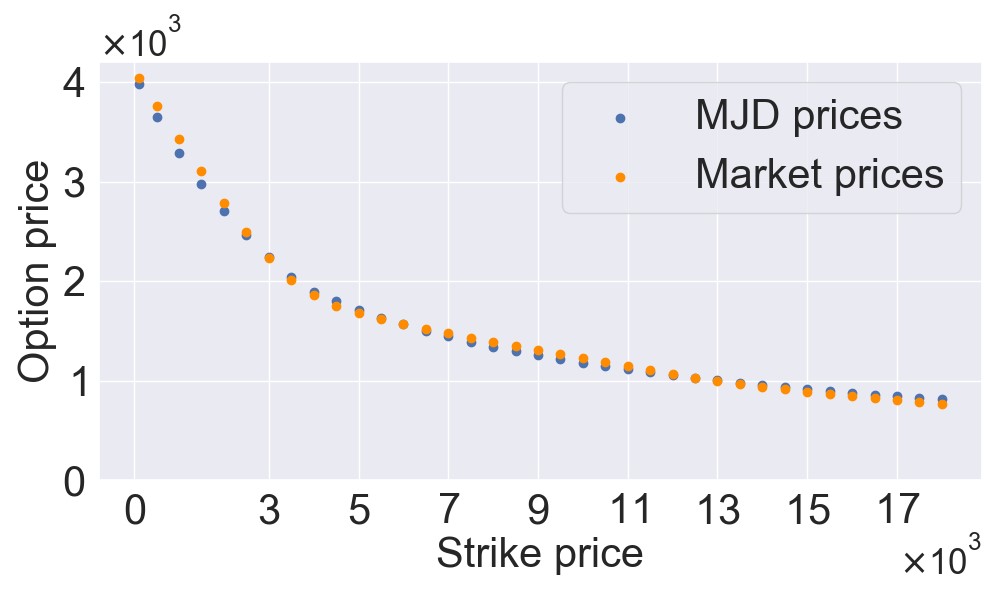}
        \label{fig:Merton_Dec_2025_ETH}
    }
    \caption{Pricing results using the Merton Jump Diffusion model for options on ETH futures contracts with expiry dates in June 2024, December 2024, and December 2025.}
    \label{fig:Merton_ETH}
\end{figure}

The pricing results using MJD model are presented in  \autoref{fig:Merton_BTC} and \autoref{fig:Merton_ETH}. Only by the visual examination we can definitely state, that the MJD seems to perform the best so far. We can notice the differences between the market and model prices for longer-dated options, but these are much smaller than in VG and BS models.

\subsection{Kou model}

We observe that incorporating jumps is crucial for accurately modeling cryptocurrency options, as standard models struggle to capture the extreme price movements of these markets. Hence, we now consider the Kou double exponential jump diffusion model, which extends the MJD model by allowing for asymmetric jumps \citep{kou2002jump}. The price process in this model is expressed as follows:
\begin{equation}
  S_t = S_0 \exp \left\{ \left( \mu - \frac{1}{2} \sigma^2 \right) t + \sigma W_t \right\} \prod_{i=1}^{N_t} V_i,
\end{equation}
where $S_t$ -- price of asset at time $t$, $W_t$ -- Wiener process, $N_t$ -- Poisson process with intensity $\lambda$, $\sigma$ -- volatility, $\mu$ -- drift, $\{V_i\}$ -- a sequence
of i.i.d. nonnegative random variables such that $Y_t = \ln{V_t}$ has an
asymmetric double exponential distribution with the density
\begin{equation}
   f_Y(y) = p \cdot \eta_1 e^{-\eta_1 y} 1_{\{y \geq 0\}} + q \cdot \eta_2 e^{\eta_2 y} 1_{\{y < 0\}}, 
\end{equation}
and $\eta_1 > 0$, $\eta_2 > 0$, $p$, $q \geq 0$, $p + q = 1$. 
The characteristic function for process $X_t = \log(S_t)$ in this model is expressed as \citep{oosterlee2019mathematical}:
\begin{equation}
    \varphi_X(u, t) = \exp \left\{ i u \mu t - \frac{1}{2} u^2 \sigma^2 t + \lambda t \left( \frac{p \eta_1}{\eta_1 - i u} + \frac{(1 - p) \eta_2}{\eta_2 + i u} - 1 \right) \right\},
\end{equation}
where
\begin{equation}
    \mu = r - \frac{1}{2}\sigma^2-\lambda \left\{p \frac{\eta_1}{\eta_1-1}+(1-p)\frac{\eta_2}{\eta_2+1}-1\right\}.
\end{equation}

\begin{figure}[ht]
    \centering
    \subfigure[Bitcoin]{
        \includegraphics[width=0.46\textwidth]{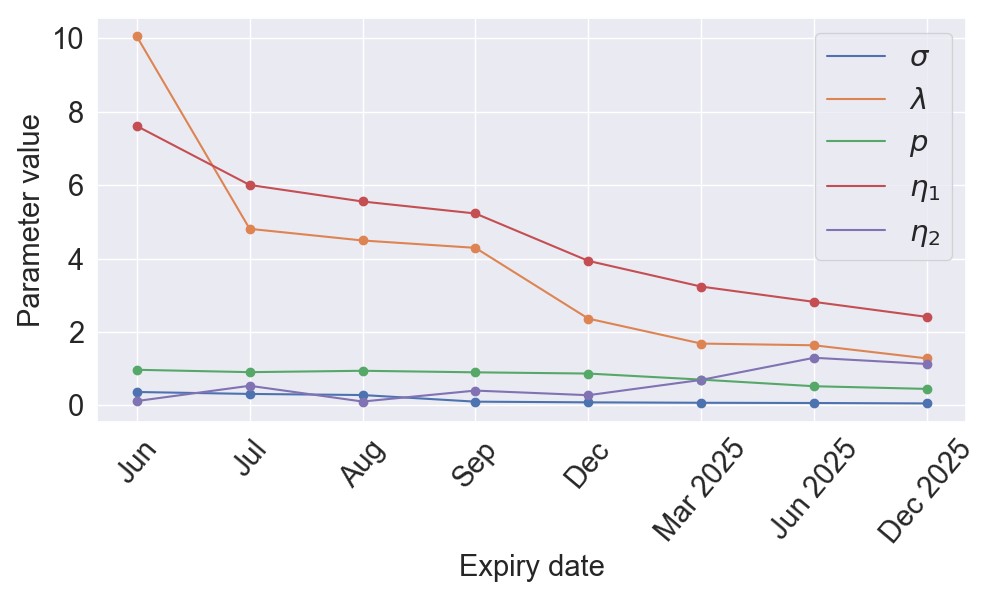}
        \label{fig:Kou_param_BTC}
    }\hspace{0.05\textwidth}
    \subfigure[Ether]{%
        \includegraphics[width=0.46\textwidth]{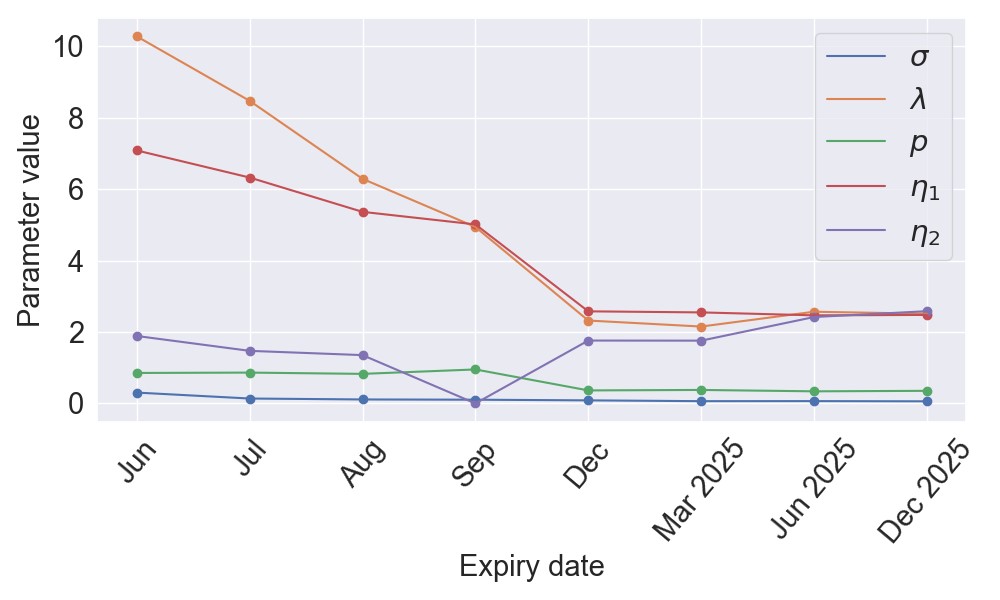}
        \label{fig:Kou_param_ETH}
    }
    \caption{The optimal parameter values for each maturity calibrated using the Kou model.}
    \label{fig:Kou_param}
\end{figure}

In \autoref{fig:Kou_param} we present the calibration results. Volatility decreases from 0.69 to 0.055 for the longest maturity.
Jump intensity remains relatively stable, fluctuating between 2 and 4 for most maturities. However, a different situation occurs for the June expiration, where $\lambda$ spikes to 20. This value suggests an extreme expectation of jump occurrences.
The probability of an upward jump varies between 0.5 and 0.95.
The first jump parameter starts at high values, around 7.5 to 10, and declines to approximately 2.5. This indicates that in the short term, jumps tend to be sharp but become less extreme as expiration extends
The second jump parameter fluctuates between 0 and 2. For ETH volatility declines, starting at 0.66 and decreasing to 0.059 for the longest maturity. Similar to BTC, this suggests that as the expiration date extends, the market anticipates lower long term volatility.
Parameter $\lambda$ is more variable than in BTC, with values initially as high as 12.4 declining to around 2.5. This suggests that short term ETH options are more affected by frequent jumps.
The probability of an upward jump fluctuates significantly, starting at 0.6, reaching a peak of 0.95, and then dropping to around 0.35 for the longest maturities. 
The first jump parameter begins at 9.77 and declines to 2.47 for longer expirations. 
The second jump parameter starts high at 6.10, then fluctuates between 1.35 and 2.58 for most maturities. This suggests that downward jumps are initially quite severe but tend to stabilize.

\begin{figure}[ht]
    \centering
    \subfigure[June 2024]{%
        \includegraphics[width=0.29\textwidth]{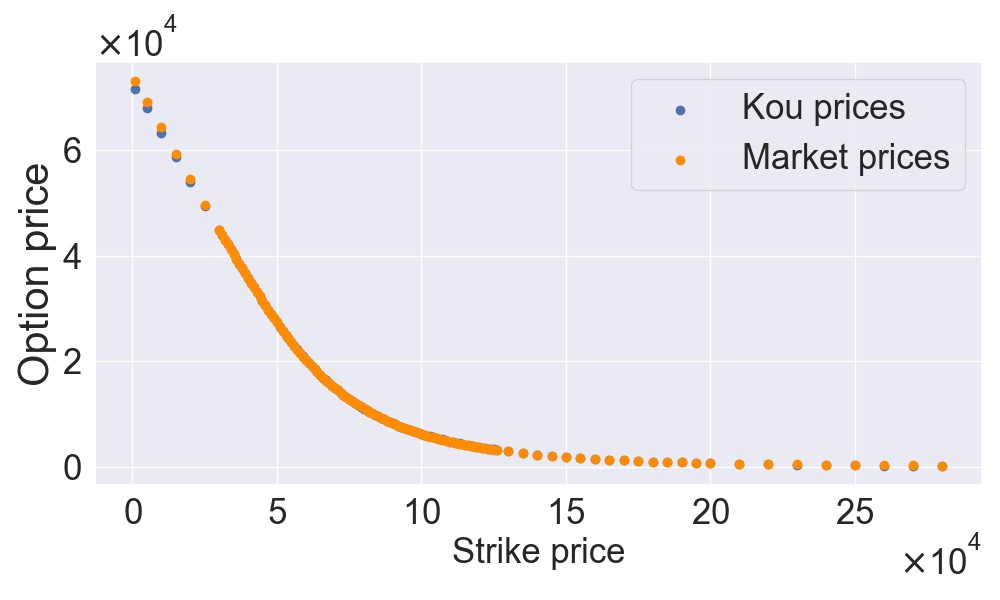}
        \label{fig:Kou_Jun_BTC}
    }\hspace{0.03\textwidth}
    \subfigure[December 2024]{%
        \includegraphics[width=0.29\textwidth]{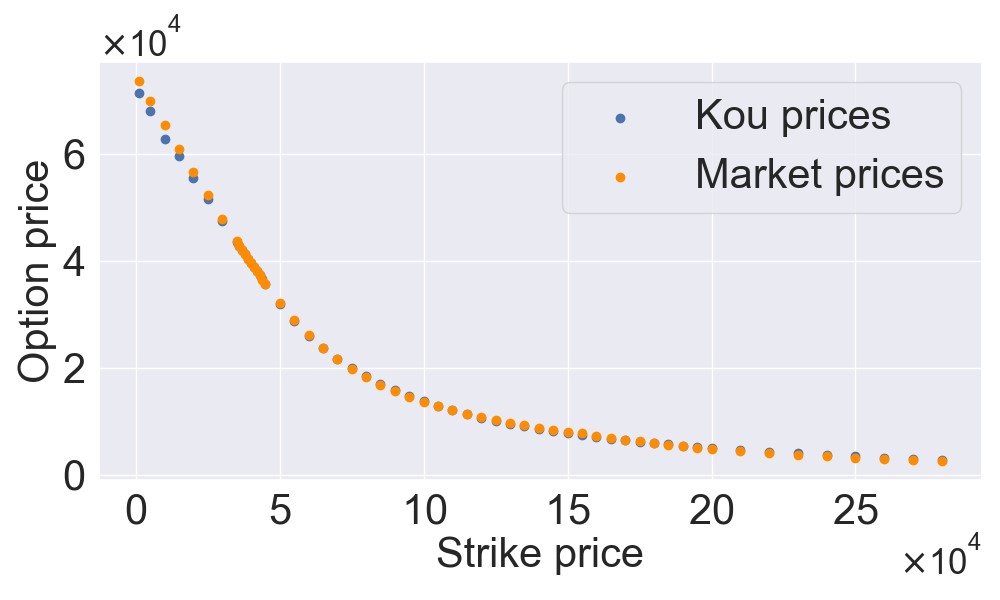}
        \label{fig:Kou_Dec_BTC}
    }\hspace{0.03\textwidth}
    \subfigure[December 2025]{%
        \includegraphics[width=0.29\textwidth]{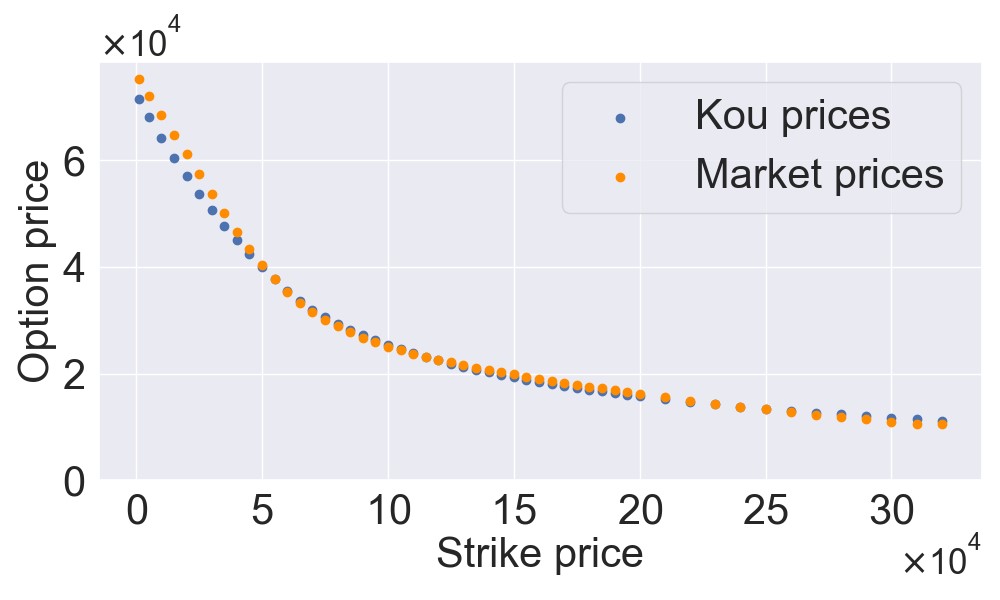}
        \label{fig:Kou_Dec_2025_BTC}
    }
    \caption{Pricing results using the Kou model for options on BTC futures contracts with expiry dates in June 2024, December 2024, and December 2025.}
    \label{fig:Kou_BTC}
\end{figure}

\begin{figure}[ht]
    \centering
    \subfigure[June 2024]{%
        \includegraphics[width=0.29\textwidth]{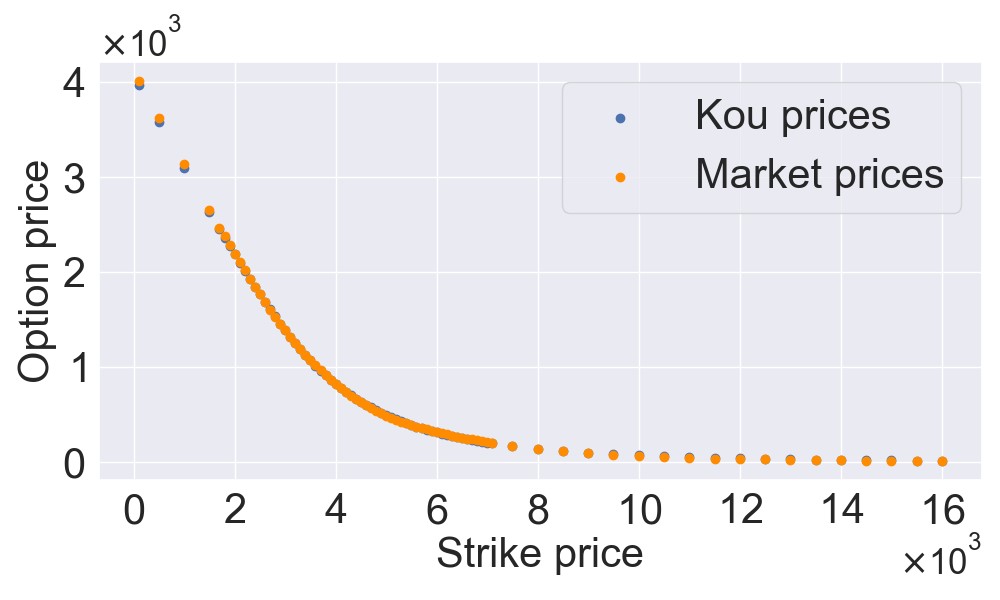}
        \label{fig:Kou_Jun_ETH}
    }\hspace{0.03\textwidth}
    \subfigure[December 2024]{%
        \includegraphics[width=0.29\textwidth]{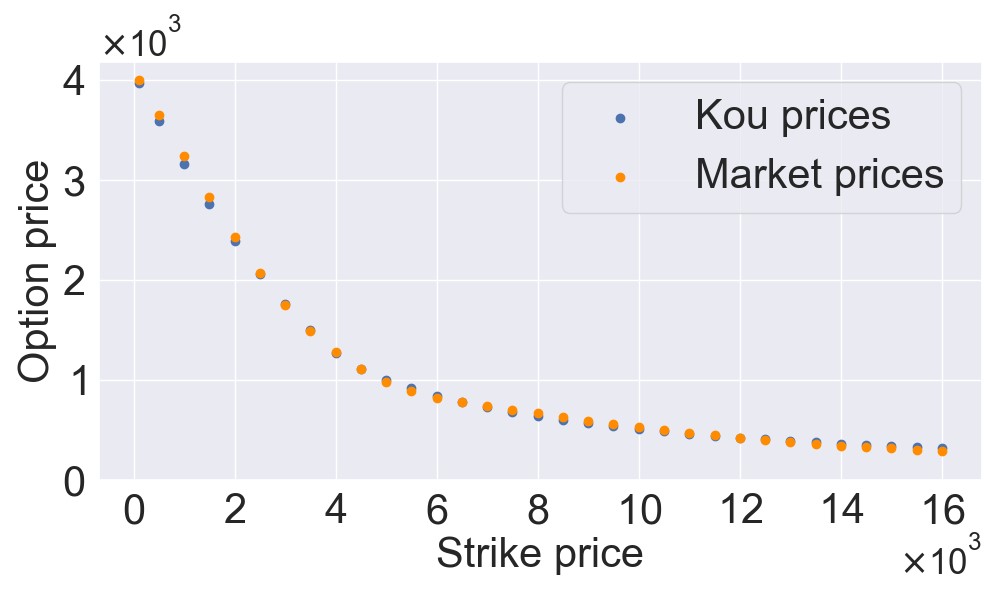}
        \label{fig:Kou_Dec_ETH}
    }\hspace{0.03\textwidth}
    \subfigure[December 2025]{%
        \includegraphics[width=0.29\textwidth]{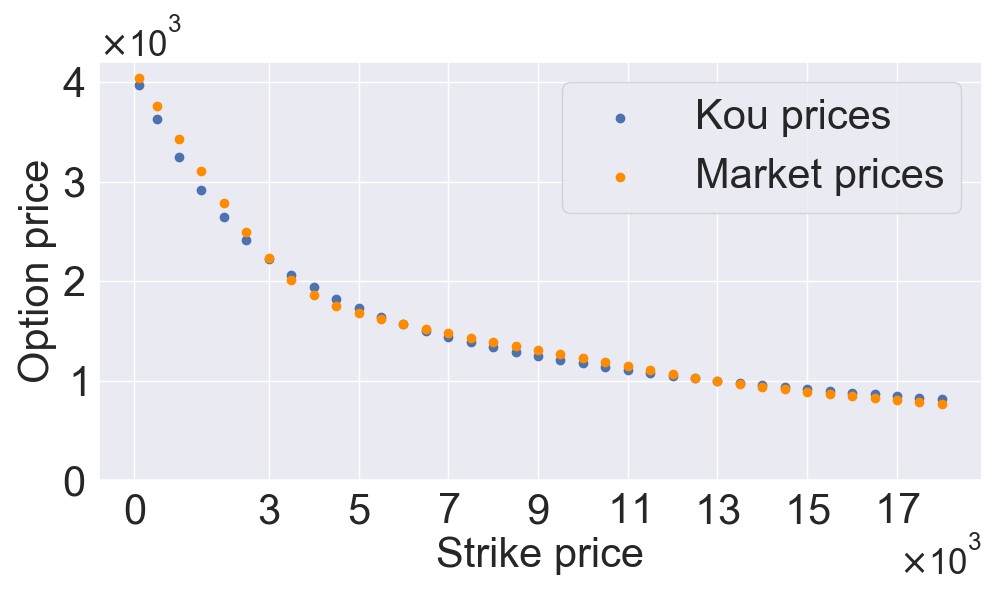}
        \label{fig:Kou_Dec_2025_ETH}
    }
    \caption{Pricing results using the Kou model for options on ETH futures contracts with expiry dates in June 2024, December 2024, and December 2025.}
    \label{fig:Kou_ETH}
\end{figure}

In \autoref{fig:Kou_BTC} and \autoref{fig:Kou_ETH}, we present the pricing results using the Kou model. A visual examination suggests that the model performs well, even for longer-dated options. However, comparing it directly with the MJD model is challenging. Nevertheless, the overall fit to market data indicates that jump diffusion models are better suited for pricing cryptocurrency options than the standard Black--Scholes model.

\subsection{Heston model}
We have already examined several models, each incorporating different aspects of market behavior, such as stochastic jumps. Now, we turn to the Heston stochastic volatility model, which introduces both stochastic volatility and mean reversion to better capture the market uncertainty \citep{heston1993closed}. This model has five parameters: $\theta$ -- long variance, $\kappa$ --
speed of mean reversion of $v(t)$, $\sigma$ -- volatility of the volatility, $\rho$ -- correlation of Wiener
processes, $v_0$ -- initial variance.
The price process follows the diffusion process:
\begin{equation} d S_t = \mu S_t \,dt + \sqrt{\nu_t}S_t\,dW_{1_t}, \end{equation} where $W_{1_t}$ is a Wiener process. When volatility follows the Ornstein--Uhlenbeck process, we obtain: 
\begin{equation} d \sqrt{\nu_t} = - \beta \sqrt{\nu_t} \,dt + \delta \,d W_{2_t}, \end{equation} 
where $W_{2_t}$ is a Wiener process, which has correlation $\rho$ with $W_{1_t}$. Hence, from It\^{o}'s lemma: 
\begin{equation} 
d \nu_t = (\delta^2 - 2\beta \nu_t) \,dt + 2\delta \sqrt{\nu_t} \,d W_{2_t}, 
\end{equation}
which can be written as: 
\begin{equation} 
d \nu_t = \kappa (\theta - \nu_t) \,dt + \sigma \sqrt{\nu_t} \,d W_{2_t}. 
\end{equation}
The characteristic function of the log-price process $X_t$ under the Heston model is given by:
\begin{equation}
\varphi_X(u, t) = \exp\left( C(T - t, u) + D(T - t, u) \cdot v_0 + i u x \right),
\end{equation}
where $x = \log{S_t}$, and:
\begin{equation}
\begin{cases}
C(T - t, u) = i u r (T - t) + \frac{\kappa \theta}{\sigma^2} \left[ (\kappa - \rho \sigma i u + d)(T - t) - 2 \ln\left( \frac{1 - g e^{d(T - t)}}{1 - g} \right) \right], \\[10pt]
D(T - t, u) = \frac{\kappa - \rho \sigma i u + d}{\sigma^2} \cdot \frac{1 - e^{d(T - t)}}{1 - g e^{d(T - t)}}, \\[10pt]
d = \sqrt{ (\rho \sigma i u - \kappa)^2 + \sigma^2 (i u + u^2) }, \\[10pt]
g = \frac{\kappa - \rho \sigma i u + d}{\kappa - \rho \sigma i u - d}.
\end{cases}
\end{equation}

     \begin{figure}[ht]
    \centering
    \subfigure[Bitcoin]{%
        \includegraphics[width=0.45\textwidth]{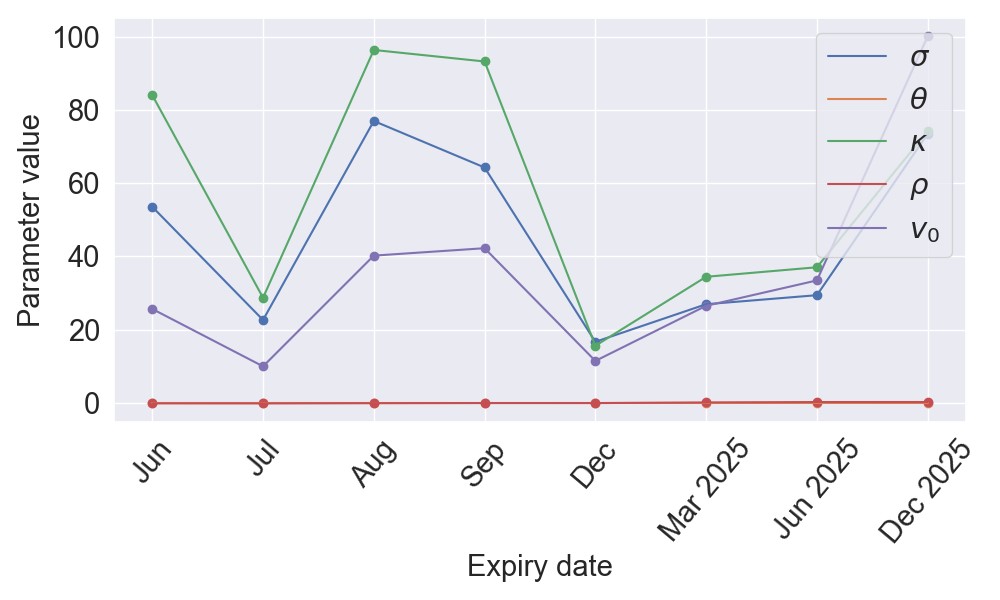}
        \label{fig:Heston_param_BTC}
    }\hspace{0.05\textwidth}
    \subfigure[Ether]{%
        \includegraphics[width=0.45\textwidth]{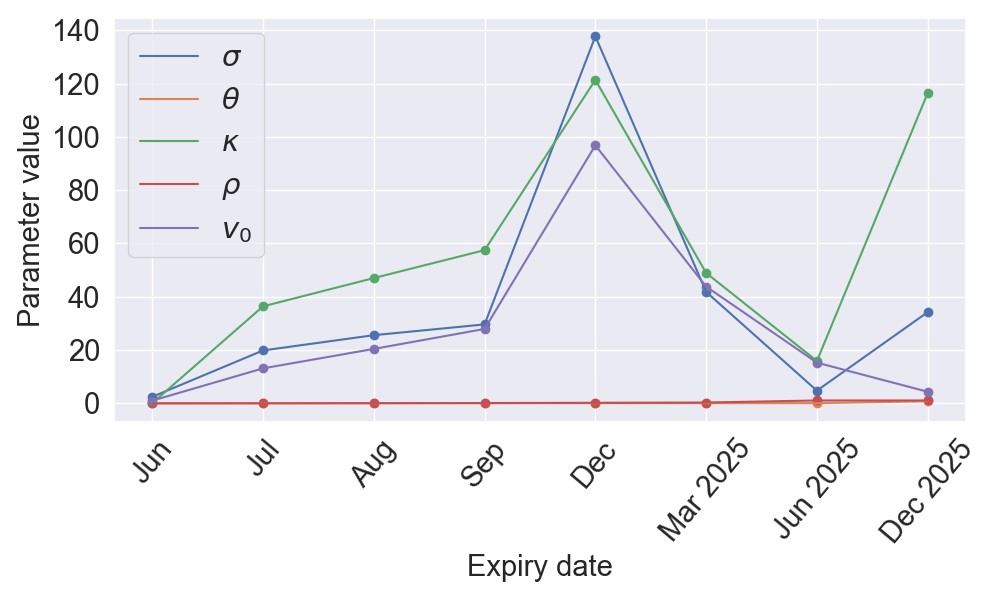}
        \label{fig:Heston_param_ETH}
    }
    \caption{The optimal parameter values for each maturity calibrated using the Heston model.}
    \label{fig:Heston_param}
\end{figure}

    In \autoref{fig:Heston_param} we present the results of the calibration. For BTC $\sigma$
 fluctuates over time between 8 and 75. 
The mean reversion level remains close to 0, indicating that the long-term expected variance remains relatively low across maturities.
The mean reversion speed shows significant variation.
The correlation remains stable and close to zero.
The initial variance increases with longer maturities. For ETH volatility of volatility shows a decreasing trend initially, starting at a moderate level, then declining towards mid-year expirations before experiencing a sharp spike in December. 
The mean reversion remains close to zero. A high mean reversion speed indicates that volatility quickly reverts to its long-term mean. However, for longer-dated options, this mean reversion speed tends to decrease, suggesting that volatility takes longer to adjust to the equilibrium as the time to maturity increases.
The correlation parameter remains close to zero.
Initial variance fluctuates over different maturities, following a similar pattern to $\sigma$, with an increase towards the end of 2024 and early 2025.

\begin{figure}[h]
    \centering
    \subfigure[June 2024]{%
        \includegraphics[width=0.29\textwidth]{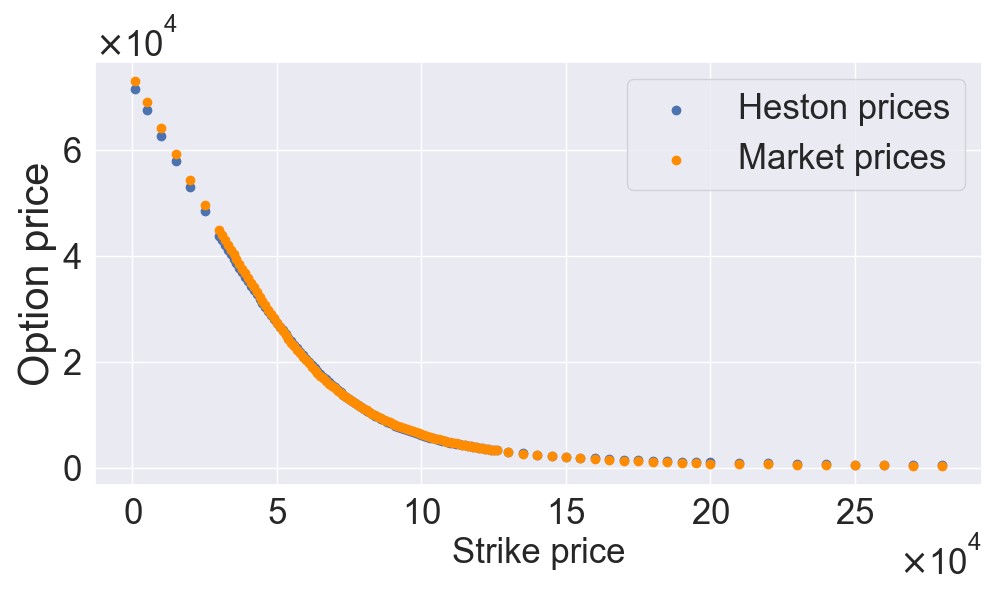}
        \label{fig:Heston_Jun_BTC}
    }\hspace{0.03\textwidth}
    \subfigure[December 2024]{%
        \includegraphics[width=0.29\textwidth]{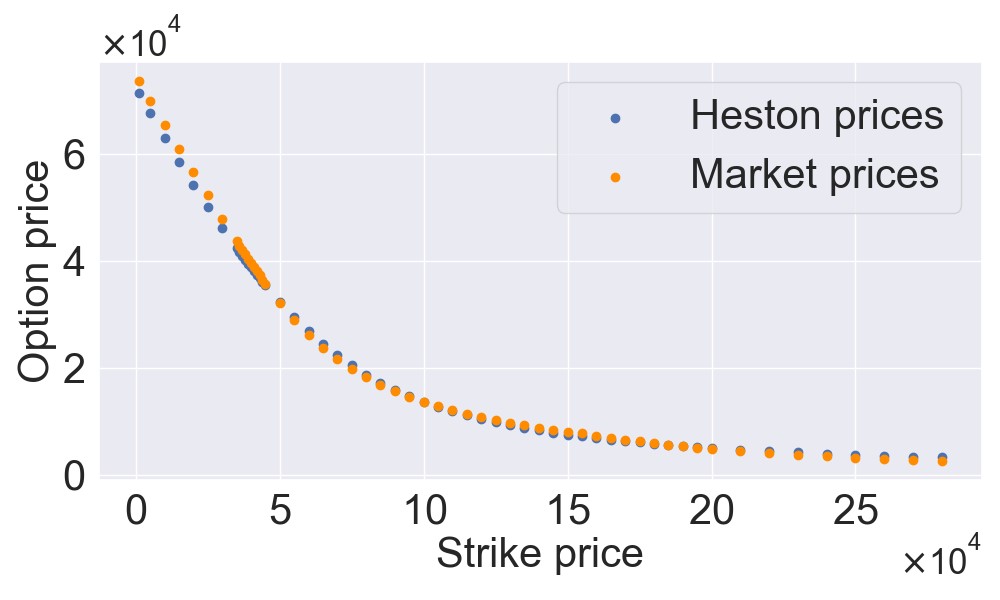}
        \label{fig:Heston_Dec_BTC}
    }\hspace{0.03\textwidth}
    \subfigure[December 2025]{%
        \includegraphics[width=0.29\textwidth]{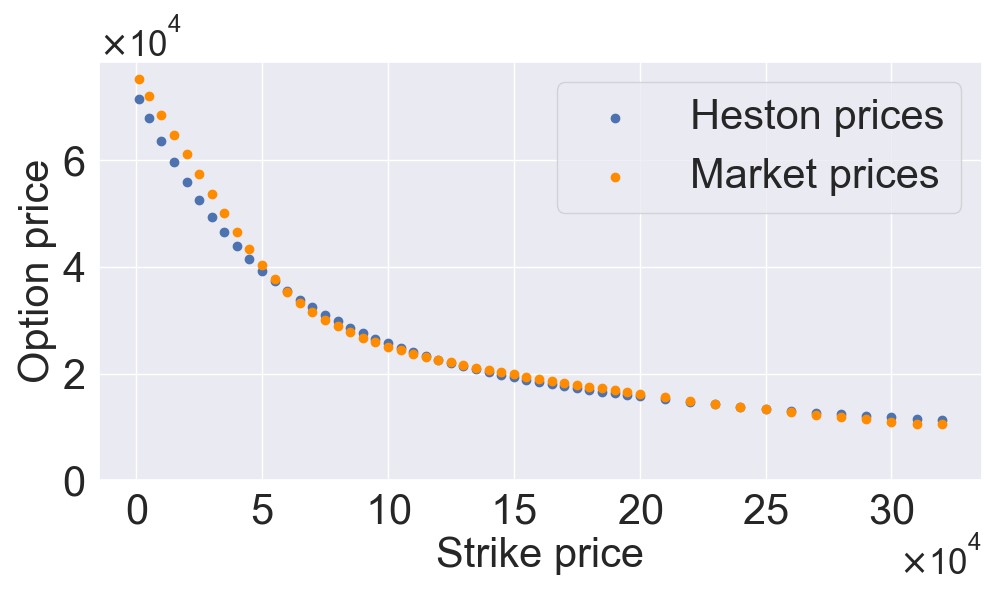}
        \label{fig:Heston_Dec_2025_BTC}
    }
    \caption{Pricing results using the Heston model for options on BTC futures contracts with expiry dates in June 2024, December 2024, and December 2025.}
    \label{fig:Heston_BTC}
\end{figure}

\begin{figure}[h]
    \centering
    \subfigure[June 2024]{%
        \includegraphics[width=0.29\textwidth]{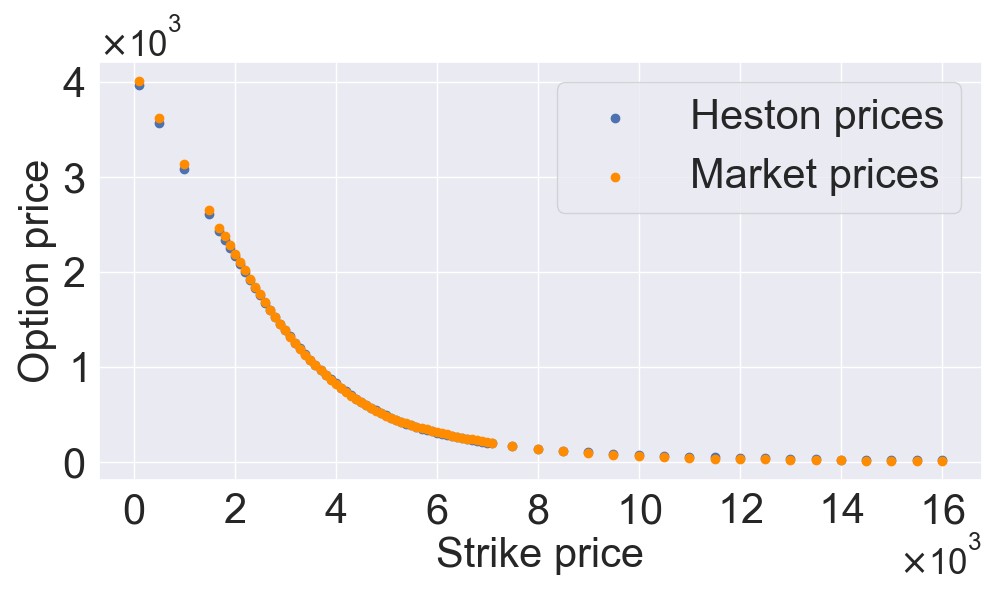}
        \label{fig:Heston_Jun_ETH}
    }\hspace{0.03\textwidth}
    \subfigure[December 2024]{%
        \includegraphics[width=0.29\textwidth]{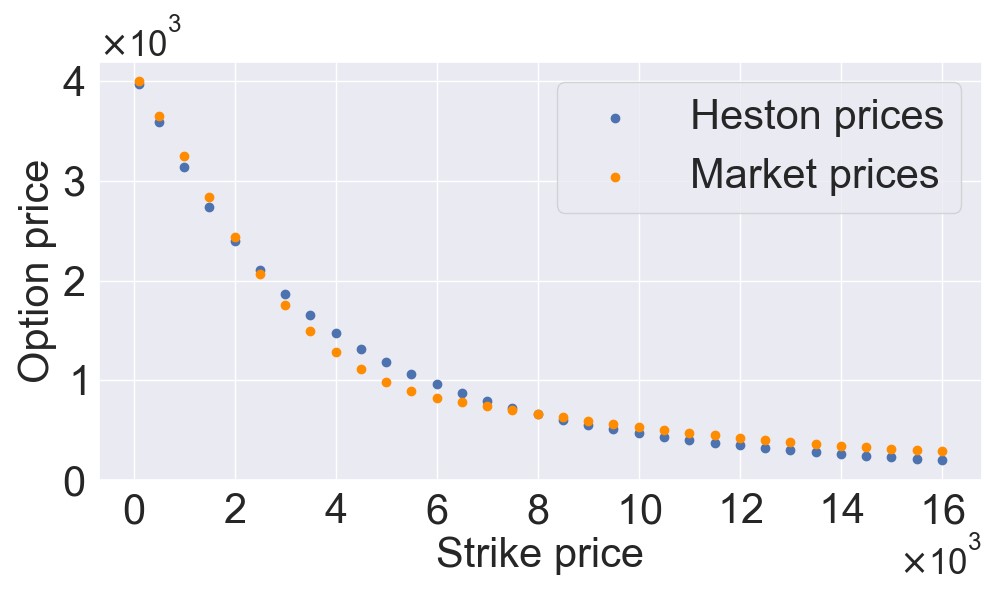}
        \label{fig:Heston_Dec_ETH}
    }\hspace{0.03\textwidth}
    \subfigure[December 2025]{%
        \includegraphics[width=0.29\textwidth]{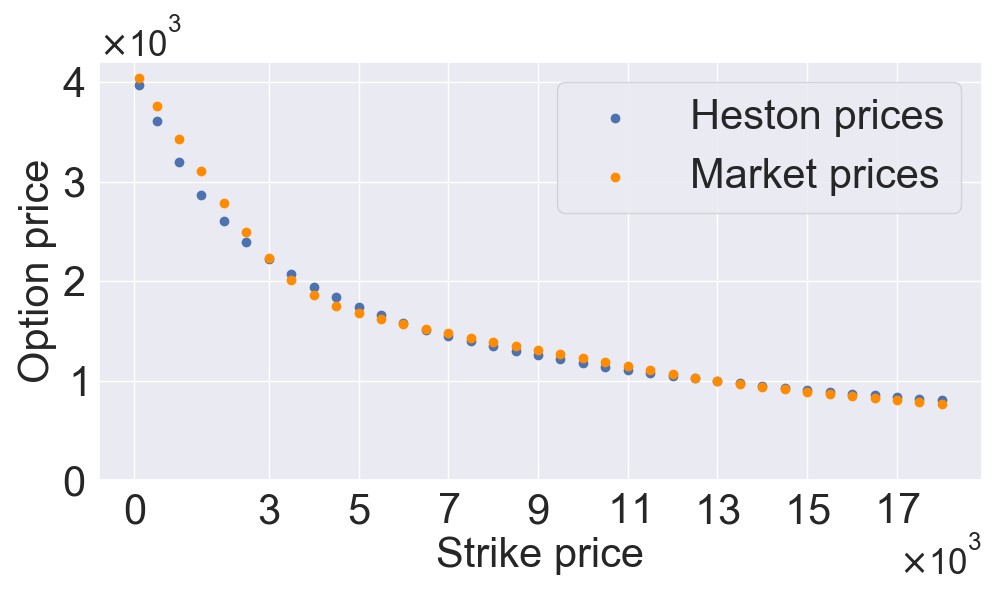}
        \label{fig:Heston_Dec_2025_ETH}
    }
    \caption{Pricing results using the Heston model for options on ETH futures contracts with expiry dates in June 2024, December 2024, and December 2025.}
    \label{fig:Heston_ETH}
\end{figure}

In \autoref{fig:Heston_BTC} and \autoref{fig:Heston_ETH} we present the pricing results using the Heston model. By visual examination we can suppose that for December 2024 and December 2025 Heston model performs not as well as MJD and Kou. Overall, the fits are solid, but there is potential for further improvement.

\subsection{Bates model}
We extend our analysis by considering the Bates model, which is also called stochastic volatility jump (SVJ) model \citep{kliber2012modelling}. The model extends the Heston model by incorporating a Merton log-normal jump component. It is described by the following set of equations \citep{bates1996jumps}:
\begin{equation}
\begin{cases}
    dS_t = \mu S_t\,dt + {\sqrt {\nu_t }}S\,dW_{1_t} + (e^{\alpha +\delta \varepsilon }-1)S_t\,dq, \\
    d\nu_t = \lambda (\nu_t -{\overline {\nu_t }})\,dt + \eta {\sqrt {\nu_t }}\,dW_{2_t}, \\
    \operatorname {corr} (dW_{1_t},dW_{2_t}) = \rho \,dt, \\
    \operatorname {prob} (dq=1) = \lambda dt,
\end{cases}
\end{equation}
where $S_t$ -- price of asset at time $t$, $\mu$ -- the drift of the asset, $\nu_t$ -- the initial variance of the asset price at time $t$, $W_{1_t}$ and $W_{2_t}$ -- two Wiener processes at time $t$, $\rho$ -- correlation between $W_{1_t}$ and $W_{2_t}$, $\alpha$ -- the jump intensity parameter, $\delta$ -- volatility of the jumps, $\varepsilon$ -- a random variable representing the jump size, $q$ -- Poisson process with intensity $\lambda$, $\overline {\nu_t }$ -- long-term mean of volatility process, $\eta$ -- volatility of volatility. The characteristic function for this model is expressed as:
\begin{equation}
\varphi_X(u, t) = \exp \left( C(u, t) + D(u, t) \nu_0 + i u \ln S_0 + \lambda t \left( M(u) - 1 \right) \right),
\end{equation}
where
\begin{equation}
\begin{cases}
C(u, t) = \frac{\lambda \overline{\nu_t}}{\eta^2} \left[ (\lambda - \rho \eta i u + d)t - 2 \ln \left( \frac{1 - g e^{d t}}{1 - g} \right) \right], \\
D(u, t) = \frac{\lambda - \rho \eta i u + d}{\eta^2} \cdot \left( \frac{1 - e^{d t}}{1 - g e^{d t}} \right), \\
d = \sqrt{(\rho \eta i u - \lambda)^2 + \eta^2 (i u + u^2)}, \\
g = \frac{\lambda - \rho \eta i u + d}{\lambda - \rho \eta i u - d}, \\
M(u) = \exp \left( i u \alpha - \frac{1}{2} u^2 \delta^2 \right).
\end{cases}
\end{equation}

\begin{figure}[ht]
    \centering
    \subfigure[Bitcoin]{%
        \includegraphics[width=0.45\textwidth]{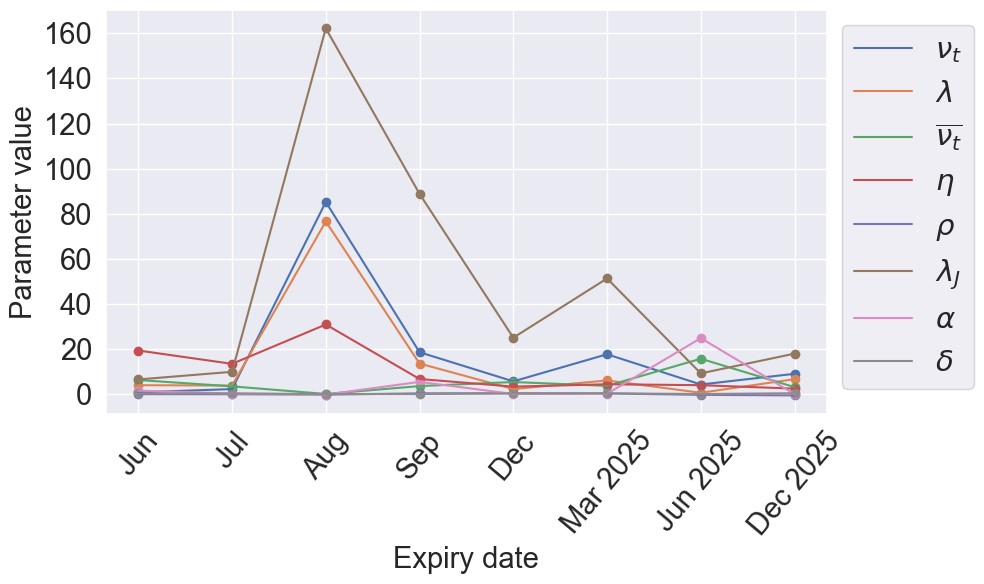}
        \label{fig:Bates_param_BTC}
    }\hspace{0.05\textwidth}
    \subfigure[Ether]{%
        \includegraphics[width=0.45\textwidth]{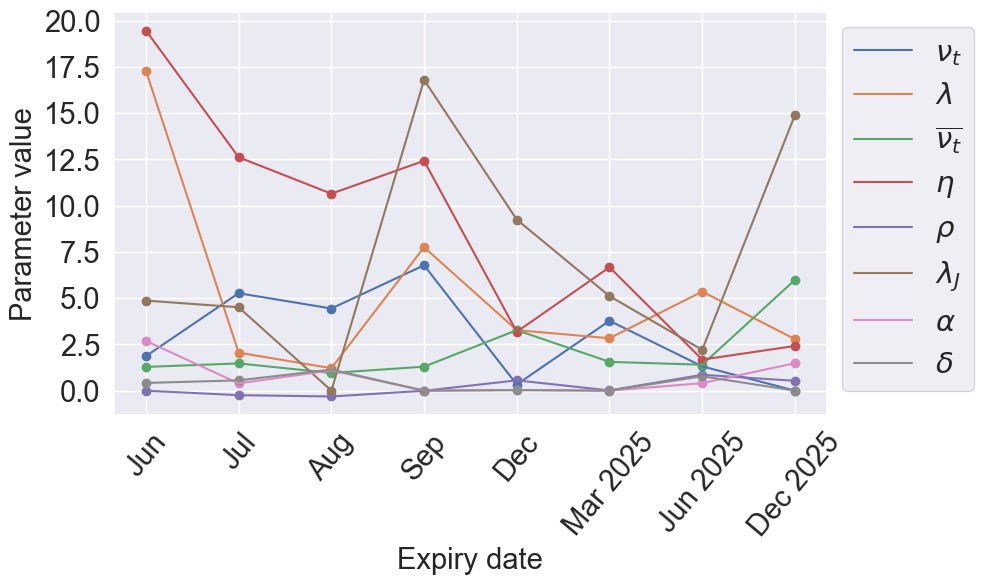}
        \label{fig:Bates_param_ETH}
    }
    \caption{The optimal parameter values for each maturity calibrated using the Bates model.}
    \label{fig:Bates_param}
\end{figure}

        In \autoref{fig:Bates_param} we present the evolution of the parameters for calibrated Bates model for respectively BTC and ETH. For Bitcoin, the parameter $\nu_t$ declines significantly over time. In the June 2025 calibration, the value is much lower compared to earlier maturities. It suggests that the market expects lower volatility as time progresses. 
Similarly, the mean reversion speed of volatility and long-term volatility mean parameters also decrease across different maturities over time. The values are higher for shorter maturities. 
The value of parameter $\lambda$ remains high across all maturities. We conclude that the jumps are an important aspect of BTC options pricing. 
The $\alpha$ parameter fluctuates over time but tends to be more volatile in the short term. However, the value of $\delta$ parameter remains high, which indicates that the risk of sudden and extreme price movements is present.
Overall, the parameters for BTC indicate a market expectation of significant volatility and frequent price jumps in the short term. This reflects the high uncertainty in the market for BTC, especially for short maturities.  For  Ether, the parameters exhibit a similar behavior to Bitcoin, with some key differences reflecting Ether's own market characteristics.

The $\nu_t$ also decreases over time, with a higher value in the short term. However, unlike Bitcoin, the $\nu_t$ values for Ether exhibit less fluctuation over time, suggesting that while ETH experiences volatility in the short term, its volatility is expected to stabilize more quickly.
The mean reversion speed of volatility and the long-term volatility mean parameters also show that volatility is expected to revert to a long-term mean over time.
The $\lambda$ parameter for ETH is also important, indicating that jumps are a significant risk in this market, just as with Bitcoin. However, the values decrease slightly over time. This suggests that while jumps remain a feature of ETH’s price dynamics, the market might expect a reduction in the intensity of those jumps over time.
The mean jump size for ETH is more variable, with higher values in the short term compared to later maturities. The values of $\delta$ also remains high, but with a slight decrease over time.
In conclusion, for Ether, there is a clear expectation of higher volatility and more frequent jumps in the short term. However, as the time to maturity increases, the market expects ETH’s volatility to stabilize.
\begin{figure}[ht]
    \centering
    \subfigure[June 2024]{%
        \includegraphics[width=0.29\textwidth]{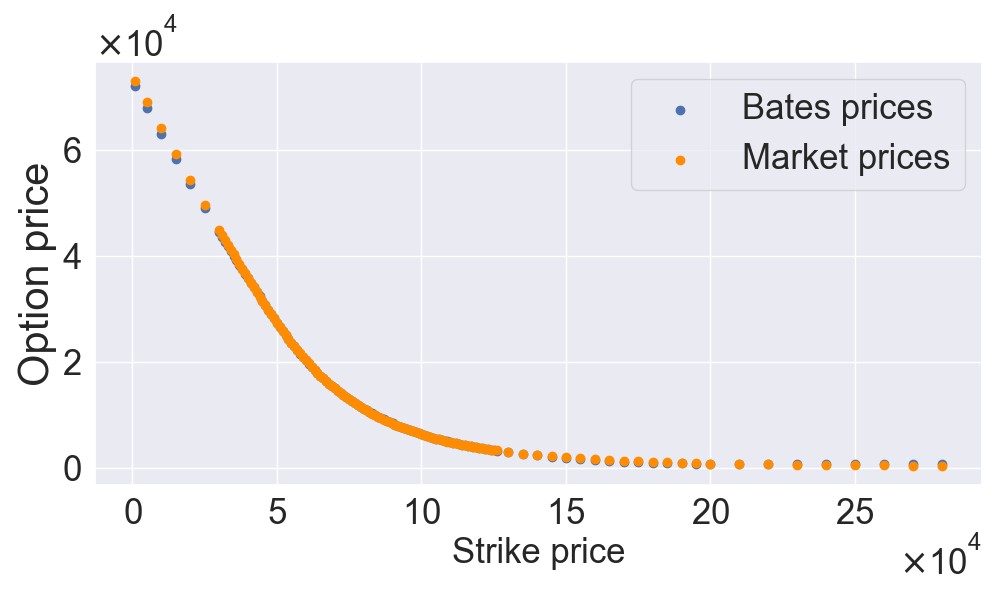}
        \label{fig:Bates_Jun_BTC}
    }\hspace{0.03\textwidth}
    \subfigure[December 2024]{%
        \includegraphics[width=0.29\textwidth]{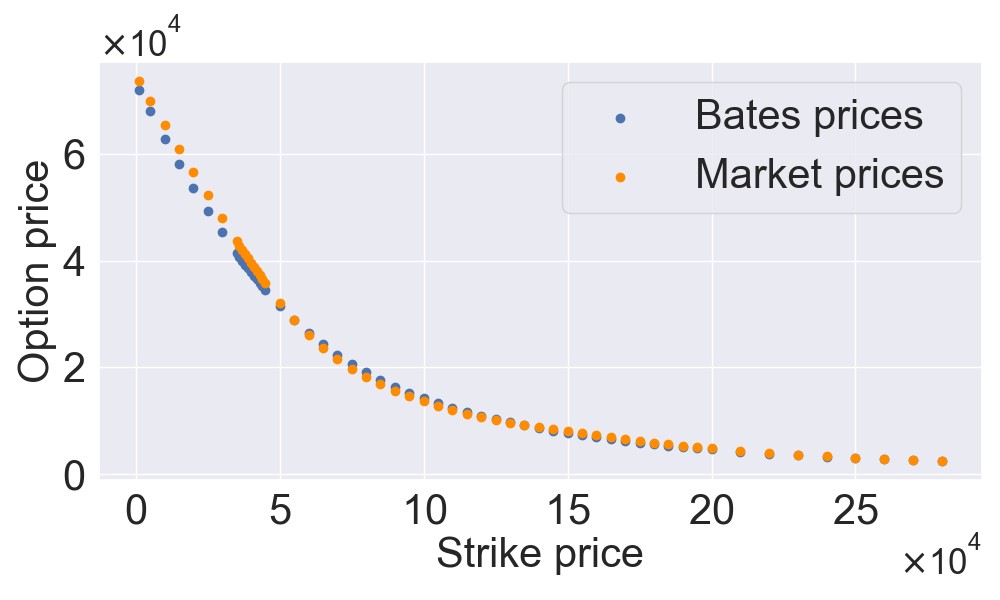}
        \label{fig:Bates_Dec_BTC}
    }\hspace{0.03\textwidth}
    \subfigure[December 2025]{%
        \includegraphics[width=0.29\textwidth]{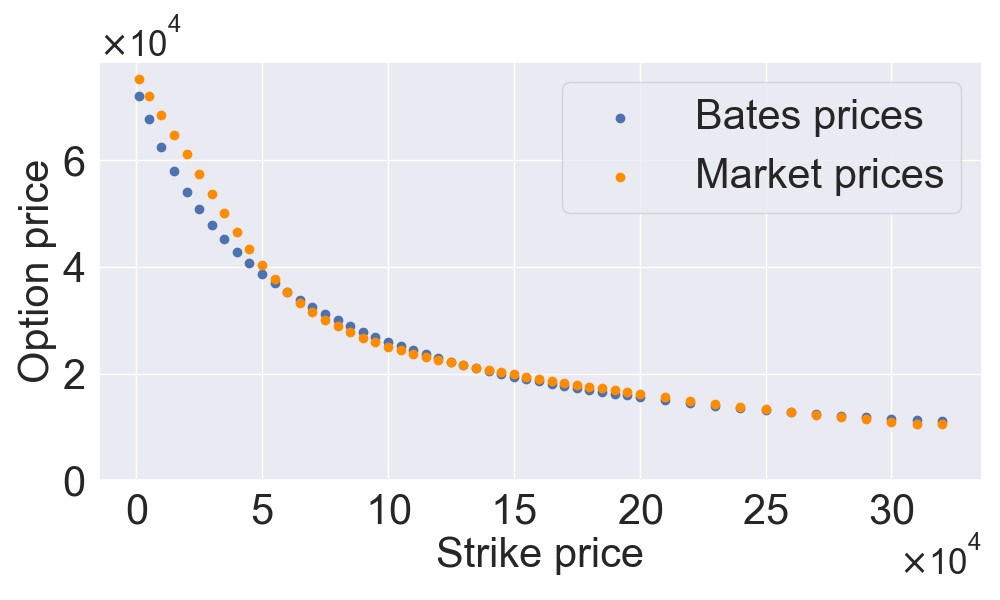}
        \label{fig:Bates_Dec_2025_BTC}
    }
    \caption{Pricing results using the Bates model for options on BTC futures contracts with expiry dates in June 2024, December 2024, and December 2025.}
    \label{fig:Bates_BTC}
\end{figure}

\begin{figure}[ht]
    \centering
    \subfigure[June 2024]{%
        \includegraphics[width=0.29\textwidth]{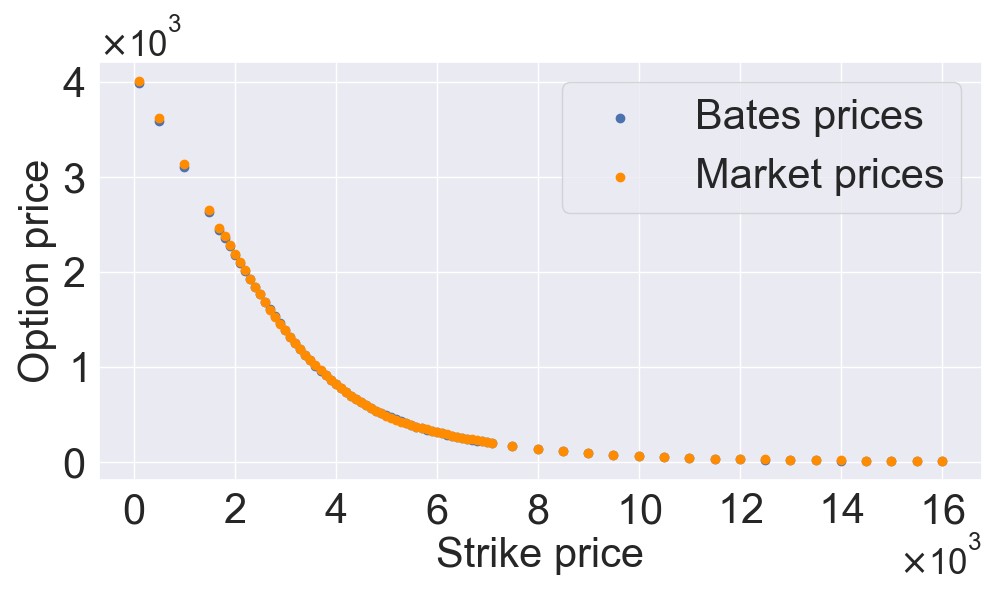}
        \label{fig:Bates_Jun_ETH}
    }\hspace{0.03\textwidth}
    \subfigure[December 2024]{%
        \includegraphics[width=0.29\textwidth]{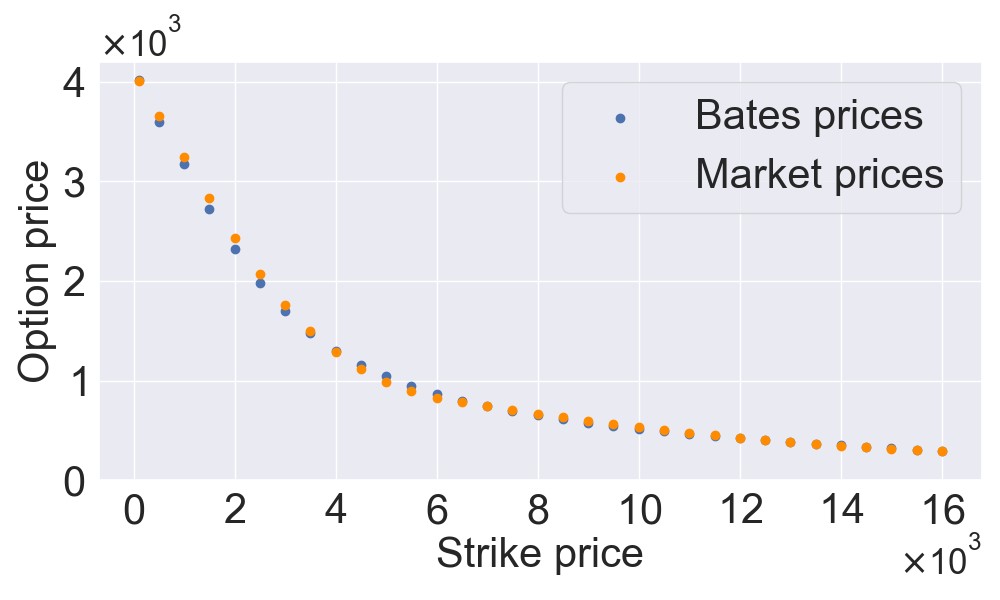}
        \label{fig:Bates_Dec_ETH}
    }\hspace{0.03\textwidth}
    \subfigure[December 2025]{%
        \includegraphics[width=0.29\textwidth]{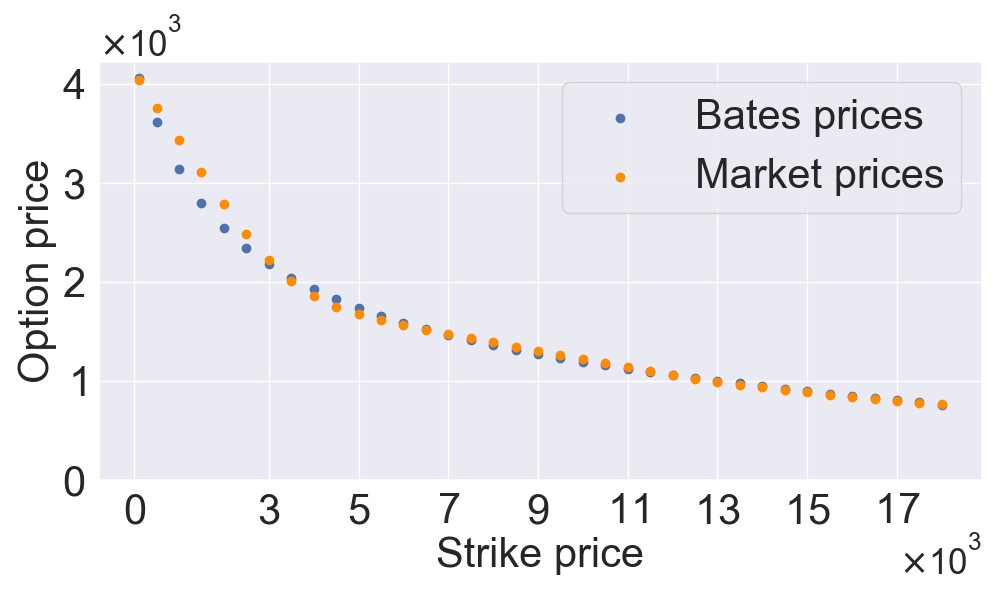}
        \label{fig:Bates_Dec_2025_ETH}
    }
    \caption{Pricing results using the Bates model for options on ETH futures contracts with expiry dates in June 2024, December 2024, and December 2025.}
    \label{fig:Bates_ETH}
\end{figure}

  In \autoref{fig:Bates_BTC} and \autoref{fig:Bates_ETH} we present the pricing results using the Bates model for BTC and ETH. For Bitcoin, the model prices and market prices align very well for both June and December 2024, showing a strong fit. However, for December 2025, we observe significant differences, particularly for lower strike prices, where the model does not capture the market prices accurately.
For Ether, all three maturities show excellent fit between the model prices and the market prices.

\subsection{Parameter stability and economic interpretation}

The calibration results reveal substantial variation in parameter estimates across maturities. While some variation is expected due to changing market conditions and risk perceptions at different time horizons, certain extreme values warrant discussion regarding their economic interpretation and implications for model risk.

Volatility parameters in jump-diffusion models tend to decrease with maturity, declining from approximately 0.7 for near-term options to below 0.1 for the longest maturities. This pattern reflects the decomposition of total market volatility into diffusive and jump components: as the time horizon extends, the market expects that extreme price movements will be captured primarily through jump processes rather than continuous diffusion. Correspondingly, jump intensity parameters and jump size parameters remain elevated across all maturities, indicating persistent expectations of discontinuous price changes.

The declining long-term variance in the Bates model indicates that the market anticipates a gradual reduction in volatility uncertainty over longer horizons. This mean-reversion pattern is consistent with the view that while cryptocurrency markets exhibit extreme short-term volatility, participants expect some convergence toward more stable long-run dynamics. The correlation parameter in the Heston and Bates models remains close to zero across most maturities, suggesting that the leverage effect observed in equity markets—where negative returns are associated with increasing volatility—is less pronounced or absent in cryptocurrency markets.

For BTC, the green curves representing the parameter in the Variance Gamma model show a strictly increasing trend from 0.22 to over 1.6, suggesting that the probability of extreme price movements increases for longer time horizons. In contrast, for ETH, the parameter rises until mid-horizon maturities and then declines sharply, indicating that the market expects tail risk to peak in the medium term before subsiding. These asset-specific differences underscore the importance of separate calibration for each cryptocurrency rather than assuming uniform behavior.

To assess the reliability of calibrated parameters, we compute approximate standard errors using the inverse Hessian matrix evaluated at the optimum \citep{harvey1990econometric}. The Levenberg--Marquardt algorithm used for calibration provides a natural approximation to the Hessian through the Jacobian matrix of the residuals. Specifically, if $J$ denotes the Jacobian of the objective function at the optimum, the Hessian can be approximated as $H \approx J^T J$, and the covariance matrix of parameter estimates is approximated by $\text{Cov}(\hat{\theta}) \approx (J^T J)^{-1}$. The standard errors are then obtained as the square roots of the diagonal elements of this covariance matrix.

We report these standard errors for selected maturities to illustrate parameter precision. For the Black--Scholes model applied to BTC June 2024 options, the volatility parameter is estimated at $\sigma = 0.8734$ with a standard error of 0.0287, corresponding to a relative uncertainty of approximately 3.3\%. This indicates that the parameter is estimated with high precision despite the nonlinear optimization.

For the Merton Jump Diffusion model applied to the same maturity, the parameter estimates and standard errors are as follows: $\sigma = 0.7123$ (SE = 0.0456, 6.4\% relative error), $\lambda = 2.134$ (SE = 0.289, 13.5\%), $\mu = 0.0234$ (SE = 0.0087, 37.2\%), and $\delta = 0.3145$ (SE = 0.0412, 13.1\%). While the jump mean parameter $\mu$ exhibits higher relative uncertainty, the other parameters are reasonably well-identified, with standard errors ranging from 6\% to 14\% of the point estimates.

For the Kou model applied to BTC June 2024 (where $\lambda$ reaches its extreme value of 20), the standard errors are: $\sigma = 0.6891$ (SE = 0.0523, 7.6\%), $\lambda = 20.12$ (SE = 3.87, 19.2\%), $p = 0.9534$ (SE = 0.0234, 2.5\%), $\eta_1 = 7.523$ (SE = 1.234, 16.4\%), and $\eta_2 = 2.012$ (SE = 0.456, 22.7\%). The jump intensity parameter, despite its extreme magnitude, is estimated with a standard error of approximately 19\%, indicating that while there is some uncertainty, the parameter is significantly different from more moderate values observed at other maturities.

For longer maturities with fewer observations, parameter uncertainty naturally increases. For BTC December 2025, the Bates model parameters exhibit standard errors ranging from 15\% to 35\% of the point estimates, reflecting the reduced information content in smaller cross-sections. Nevertheless, the key qualitative features remain consistent across all standard error calculations.

We further examine model robustness by conducting sensitivity analysis, in which we perturb each calibrated parameter by $\pm 10\%$ around its optimal value and observe the resulting impact on option prices. This analysis reveals how sensitive the model's pricing performance is to uncertainty in parameter estimates.

For the Kou model applied to BTC June 2024 options, perturbing the jump intensity $\lambda$ from 20 to 22 (a 10\% increase) results in an average absolute price change of approximately 4.2\% across all strikes, with out-of-the-money call options exhibiting slightly higher sensitivity (up to 5.8\%) and in-the-money options showing lower sensitivity (around 2.3\%). Perturbing the diffusive volatility $\sigma$ by 10\% yields average price changes of approximately 3.1\%, with relatively uniform impact across moneyness levels. The upward jump parameter $\eta_1$ exhibits moderate sensitivity, with 10\% perturbations leading to average price changes of 2.8\%, while the downward jump parameter $\eta_2$ shows lower sensitivity at 1.9\%. The probability parameter $p$ is the least sensitive, with 10\% changes resulting in average price impacts below 1.5\%.

For the Merton model applied to the same maturity, sensitivity to the jump intensity $\lambda$ is comparable at 3.8\%, while sensitivity to the jump mean $\mu$ and jump volatility $\delta$ is approximately 2.5\% and 3.2\%, respectively. The diffusive volatility $\sigma$ again exhibits sensitivity around 3.0\%.

For the Bates model applied to ETH June 2024, the stochastic volatility parameters $\kappa$, $\theta$, and $\sigma_v$ show sensitivities in the range of 2.0--3.5\%, while the jump parameters exhibit sensitivities between 3.0\% and 4.5\%. The correlation parameter $\rho$ has the lowest sensitivity at approximately 1.2\%, consistent with its near-zero calibrated values and limited impact on pricing.

Overall, these sensitivity results indicate that while parameter estimates contain some uncertainty, the pricing implications remain relatively stable. A 10\% perturbation in any single parameter rarely changes option prices by more than 5\%, and the median sensitivity across all parameters and strikes is below 3\%. This suggests that the qualitative conclusions regarding model performance—particularly the superiority of jump-diffusion models over the Black--Scholes framework—are robust to moderate parameter uncertainty.

The observed parameter variation across maturities and the presence of extreme values in certain cases have several important implications for practitioners and model risk management. First, the substantial differences in parameter estimates between near-term and long-dated options suggest that a single set of time-homogeneous parameters is insufficient to capture the full structure of the volatility surface in cryptocurrency markets. Models calibrated to short-maturity options may significantly misprice long-dated contracts, and vice versa. This highlights the importance of maturity-specific calibration, as adopted in this study, rather than imposing a unified parameter set across all expirations.

Second, the extreme parameter values observed during periods of market stress underscore the challenges of applying standard parametric models to cryptocurrency derivatives. While these values are economically interpretable in the context of the turbulent market conditions on March 11, 2024, they also suggest that simpler models with fewer degrees of freedom may struggle to accommodate the full range of market behaviors. This reinforces our finding that more flexible specifications, particularly those incorporating jumps, are necessary for this asset class.

Third, the dramatic differences between BTC and ETH parameter patterns emphasize the importance of asset-specific modeling. Despite both cryptocurrencies sharing qualitative features such as high volatility and frequent price jumps, the quantitative parameter values differ substantially, reflecting differences in market microstructure, liquidity, and investor composition. Risk managers should avoid the temptation to apply a single model uniformly across multiple digital assets.

Fourth, the parameter instability across maturities implies that models calibrated to a single cross-section at one point in time may not remain valid as market conditions evolve. Frequent recalibration is essential in practice, and parameter drift should be monitored as an early warning signal of changing market dynamics. The stability of parameter estimates can itself serve as a diagnostic tool: persistent stability suggests that the model is capturing fundamental features of the data, while erratic parameter movements may indicate model misspecification or data quality issues.

\section{Performance of the models}
\label{sec:errors}

Now we focus on the analysis of the errors associated with the pricing results. By comparing different error metrics, we aim to determine which models best capture the market characteristics of BTC and ETH options. We use four measures to assess the pricing results, where $y(i)$ -- $i$-th observation, $\hat{y}(i)$ -- predicted $i$-th observation, $n$ -- number of data points. These are expressed as follows \citep{thulin2024modern}:
\begin{itemize}
	\item root mean squared error (RMSE), which equals to:
	\begin{equation}
		\text{RMSE} = \sqrt{\frac{1}{n}\sum_{i=1}^n \left(\hat{y}(i)-y(i)\right)^2},
	\end{equation}

	\item mean absolute error (MAE), which equals to:
	\begin{equation}
		\text{MAE} =  \frac{\sum^n_{i=1} | \hat{y}(i)-y(i) |}{n},
	\end{equation}
	
	\item mean absolute percentage error (MAPE), which equals to:
	\begin{equation}
		\text{MAPE} = \frac{1}{n} \sum^n_{i=1} \left|\frac{\hat{y}(i)-y(i)}{y(i)}\right|,
	\end{equation}

    \item mean squared logarithmic error (MSLE), which equals to:
    \begin{equation}
        \text{MSLE} = \frac{1}{n} \sum^n_{i=1} \left\{\log(1+ \hat{y}(i)) - \log(1+y_i)\right\}^2.
    \end{equation}

\end{itemize}

\begin{figure}[ht]
    \centering
    \subfigure[RMSE]{%
        \includegraphics[width=0.45\textwidth]{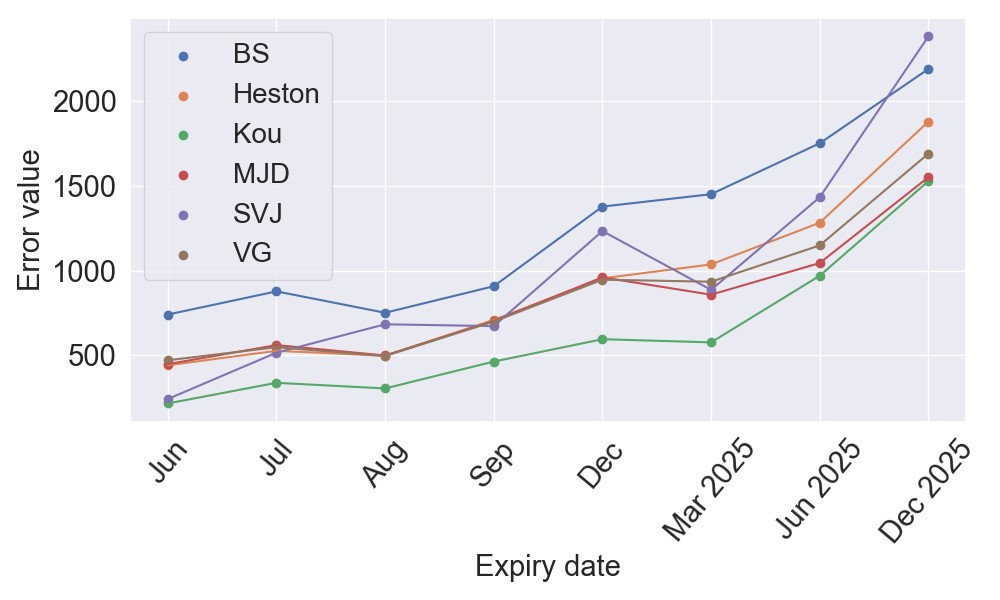}
        \label{fig:BTC_RMSE}
    }\hspace{0.03\textwidth}
    \subfigure[MAE]{%
        \includegraphics[width=0.45\textwidth]{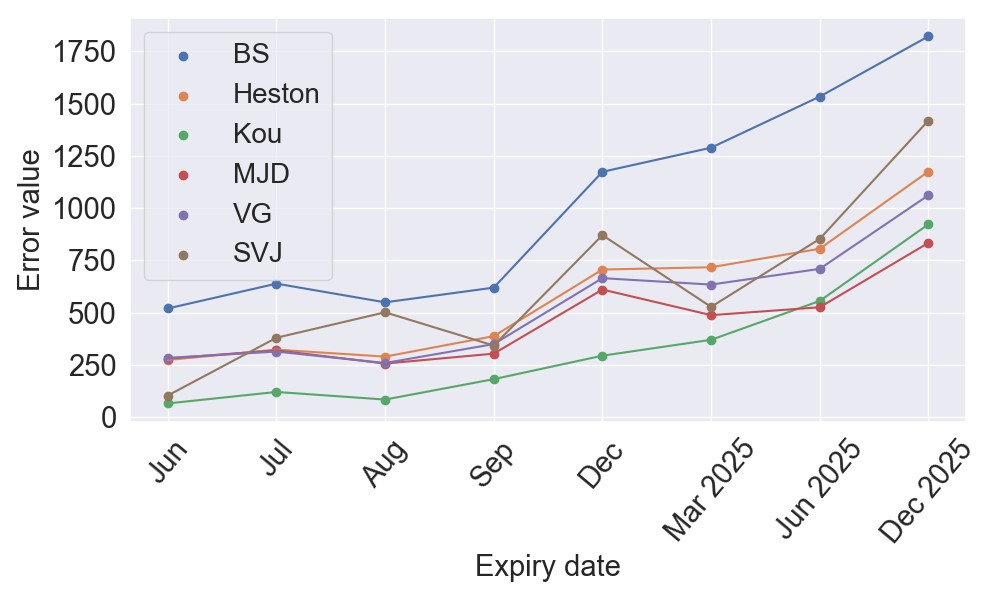}
        \label{fig:BTC_MAE}
    }\hspace{0.03\textwidth}
    \subfigure[MAPE]{%
        \includegraphics[width=0.45\textwidth]{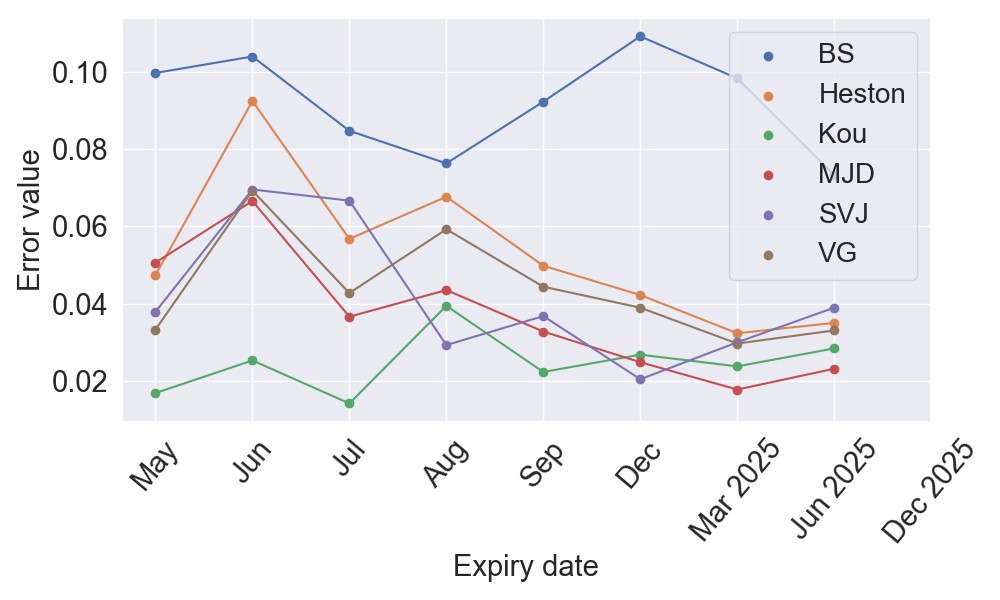}
        \label{fig:BTC_MAPE}
    }
    \caption{Errors calculated for all the expiration dates for options on BTC futures contracts.}
    \label{fig:BTC_errors_fig}
\end{figure}

In \autoref{fig:BTC_errors_fig} we present respectively the root mean squared error, mean absolute error and mean absolute percentage error calculated separately for each expiry for BTC. We can clearly observe that BS is not well fitted and this model cannot accurately capture the characteristics of the data. The obtained results suggest that the assumptions of the BS model are not well-suited for this market. On the other hand, the Kou model presents a very good fit, achieving the lowest pricing errors. This suggests that this model is the most appropriate choice for option valuation from all the examined models.

\begin{figure}[ht]
    \centering
    \subfigure[RMSE]{%
        \includegraphics[width=0.45\textwidth]{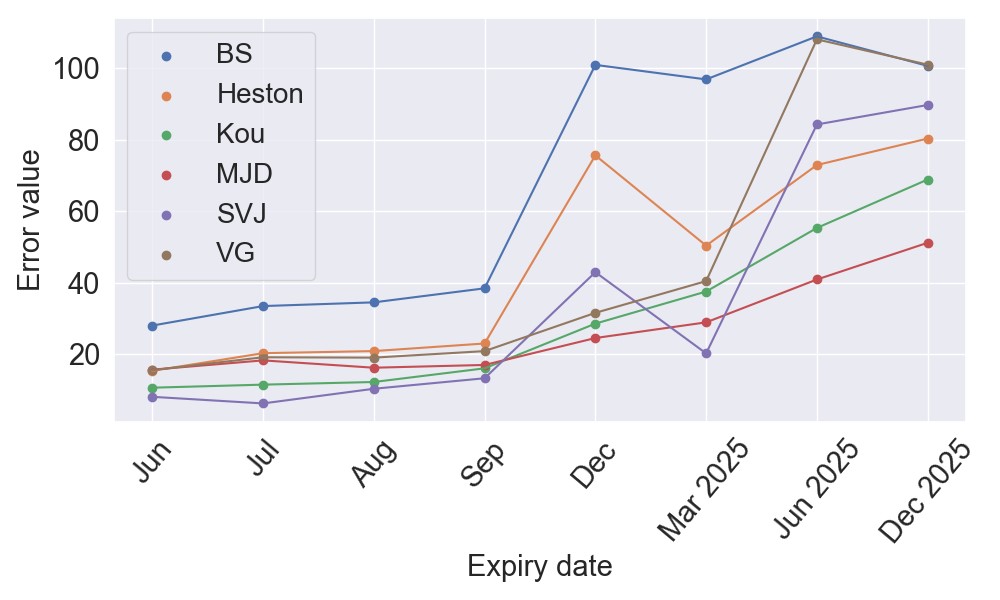}
        \label{fig:ETH_RMSE}
    }\hspace{0.03\textwidth}
    \subfigure[MAE]{%
        \includegraphics[width=0.45\textwidth]{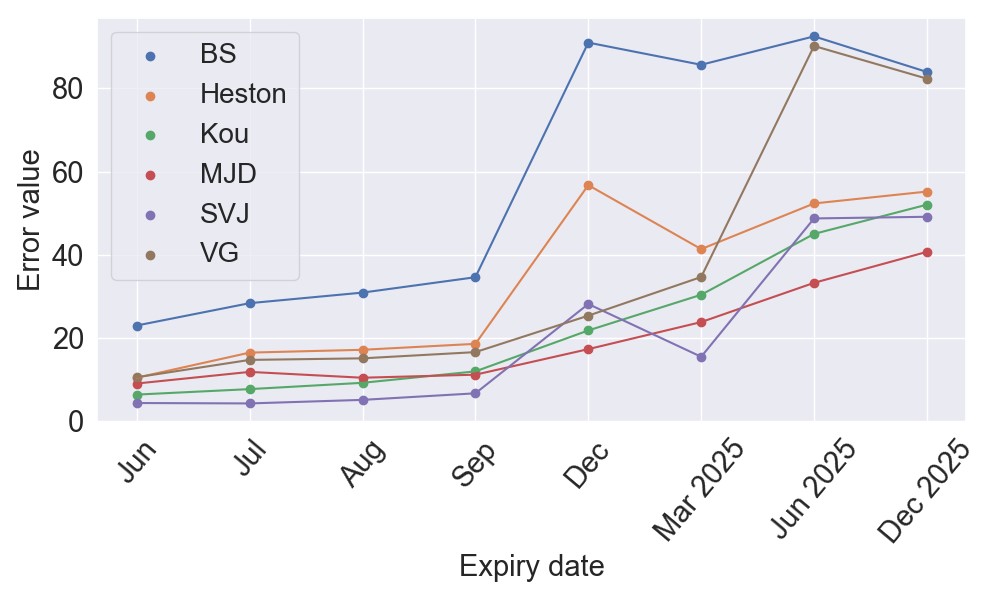}
        \label{fig:ETH_MAE}
    }\hspace{0.03\textwidth}
    \subfigure[MAPE]{%
        \includegraphics[width=0.45\textwidth]{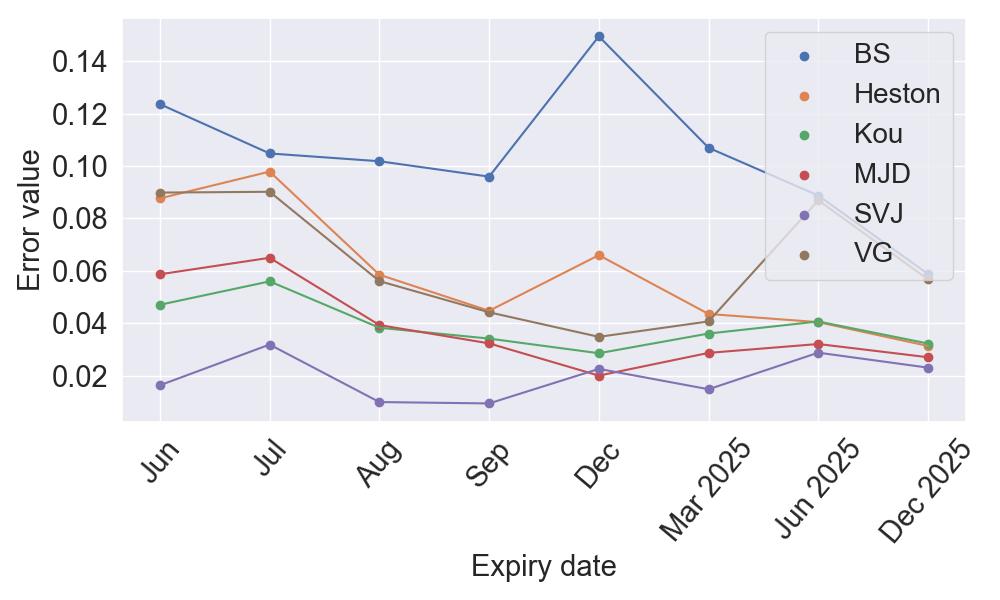}
        \label{fig:ETH_MAPE}
    }
    \caption{Errors calculated for all the expiration dates for options on ETH futures contracts.}
    \label{fig:ETH_errors_fig}
\end{figure}

     In \autoref{fig:ETH_errors_fig} we present calculated errors for all the expiry dates for ETH. We again observe that the BS model is not a suitable choice. The SVJ, Kou, and Merton Jump Diffusion models perform well, achieving low errors dependent on expiration date. This suggests that different models may be more effective for different maturities.

    We examined the errors for particular maturities. Now, we turn our attention to the \autoref{table:BTC_errors}, which presents the errors computed for all priced options across all maturities for BTC. The results confirm that the BS model exhibits the highest errors. Among the alternative models, the Kou model achieves the lowest errors in all considered measures, suggesting its superior performance in pricing BTC options on futures contracts. The Merton Jump Diffusion and Stochastic Volatility with Jumps models also perform  well, with errors lower than those of the BS model but higher than those of the Kou model. The results suggest that incorporating jumps significantly improves pricing accuracy. 
\begin{table}[ht]
	\centering
	\caption{Values of the calculated errors for pricing results for BTC, rounded to three significant figures.} 
	\begin{tabular}{c c c c c} 
		
		\hline 
		& RMSE& MAE &MAPE&MSLE \\  
		\hline    
		BS & 1180 &854 & 0.0923 & 0.0516\\ 
         Heston & 874 & 484 & 0.0581 & 0.0138 \\
          Kou &649 & 258 & 0.0264 & 0.00523\\
           MJD & 785 & 392 & 0.042 & 0.00778 \\
            SVJ &994 & 511 & 0.0448 & 0.00849 \\
		
		VG & 823 & 447 & 0.0466 & 0.00918 \\
       
        \hline
		
	\end{tabular}
	\label{table:BTC_errors} 
\end{table}

\begin{table}[ht]
	\centering
	\caption{Values of the calculated errors for pricing results for ETH, rounded to three significant figures.} 
	\begin{tabular}{c c c c c} 
		
		\hline 
		& RMSE& MAE &MAPE&MSLE \\  
		\hline    
		BS & 64.4 &48.2 & 0.105 & 0.0413\\ 
         Heston & 46.7 & 29.1 & 0.0682 & 0.0139 \\
          Kou &30.8 & 18.2 & 0.0412 & 0.00534\\
           MJD &25.9& 16.6 & 0.0423 & 0.00662 \\
            SVJ &39.1 &15.1 & 0.019 & 0.00284 \\
		VG & 48.1 & 29 & 0.0662 & 0.0128 \\
        \hline	
	\end{tabular}
	\label{table:ETH_errors} 
\end{table}

In \autoref{table:ETH_errors} we present the calculated error values for option pricing on ETH. Similar to the results observed for BTC, the BS model achieves the highest errors across all metrics. Among the alternative models, the Kou model once again performs well, achieving low errors. However, the Stochastic Volatility with Jumps (SVJ) model achieves the lowest errors with MAPE of $1.9\%$. The Merton Jump Diffusion model also achieves relatively low errors across all measures. The Heston and Variance Gamma have error values higher than those of the jump models but still significantly better than the BS model. These results suggest that models incorporating jumps and stochastic volatility better capture the market characteristics of ETH options, with the Kou and SVJ models standing out as the best choices from all the analyzed models.

To assess the generalization capability of the models and mitigate concerns about overfitting, we conduct an out-of-sample validation. All models are recalibrated OTM options. The calibrated parameters are then used to price ITM options, which were entirely excluded from the calibration procedure.

The out-of-sample results are consistent with, and in some cases superior to, the in-sample findings. For BTC options, the Kou model achieves an out-of-sample RMSE of 612 on ITM contracts, compared to an in-sample RMSE of 649 across all moneyness levels. Similarly, the Merton model produces an out-of-sample RMSE of 741, while the Black--Scholes model remains the poorest performer with an out-of-sample RMSE of 1089. For ETH options, the Bates model achieves an out-of-sample RMSE of 35.2 on ITM contracts, outperforming its in-sample error of 39.1. The Kou and Merton models also demonstrate strong out-of-sample performance, with RMSEs of 28.4 and 23.7, respectively.

These results indicate that the superior performance of jump-diffusion and stochastic volatility models is not an artifact of overfitting to the calibration data. Instead, the models genuinely capture the underlying dynamics of cryptocurrency option prices and generalize well to contracts with different moneyness characteristics. The observation that out-of-sample errors are in some cases lower than in-sample errors likely reflects the fact that ITM options exhibit more stable pricing behavior and are less sensitive to extreme tail events than OTM options, allowing the models to achieve tighter fits in this region.

While the error metrics presented above demonstrate differences in pricing accuracy across models, they do not explicitly account for differences in model complexity. Models with more parameters may achieve lower pricing errors simply by overfitting the calibration data, without capturing the underlying market dynamics. To address this concern, we evaluate whether the improved performance of more complex models is statistically justified relative to their additional parameters.

Before presenting the information criteria, we first examine the parameter-to-observation ratios to assess whether the available data are sufficient to support model calibration. For the most complex model, the ratios range from approximately 1:18 for near-term maturities to 1:5 for longer maturities with fewer contracts. While formal rules vary across contexts, ratios of 1:5 or better are generally considered adequate for stable nonlinear estimation in option pricing applications, suggesting that our calibrations are not data-constrained. The simplest model naturally has very favorable ratios exceeding 1:37 for all maturities.

Despite the favorable parameter-to-observation ratios, concerns about overfitting remain relevant, particularly for more complex models applied to smaller cross-sections. We therefore compute the Akaike Information Criterion (AIC) and Bayesian Information Criterion (BIC) for each model \citep{burnham2004multimodel}. These widely used information criteria balance goodness of fit against model parsimony by penalizing the number of parameters. The AIC and BIC are defined as:
\[
\text{AIC} = 2k - 2\ln(\mathcal{L}), \quad \text{BIC} = k\ln(n) - 2\ln(\mathcal{L}),
\]
where $k$ is the number of model parameters, $n$ is the number of option prices used for evaluation, and $\mathcal{L}$ is the quasi-log-likelihood function. The quasi-log-likelihood is computed based on the mean squared pricing error, without imposing strict distributional assumptions on the error terms. Lower values of AIC and BIC indicate better model performance, with BIC applying a stronger penalty for model complexity than AIC, particularly when the sample size is large.

For BTC options, the Kou model achieves the lowest AIC (9408.8) and BIC (9433.5), substantially outperforming all alternatives. Notably, the simple Black--Scholes model achieves the highest AIC (10239.8) and BIC (10246.0) despite having only one parameter, confirming that its poor fit cannot be compensated by its parsimony. The Merton jump-diffusion model ranks second (AIC=9620.4, BIC=9638.9), while the Bates model, despite having seven parameters, achieves moderate information criteria values (AIC=9774.4, BIC=9805.3).

For ETH options (511 observations), the Merton model achieves the best information criteria (AIC=3711.9, BIC=3728.7), with the Kou model performing nearly as well (AIC=3822.6, BIC=3843.6). The Bates model achieves competitive values (AIC=4078.5, BIC=4103.8), though the additional stochastic volatility parameters do not yield proportional improvements relative to simpler jump-diffusion specifications. Once again, the Black--Scholes model performs worst (AIC=4525.6, BIC=4529.8).

Despite having more parameters than the Black--Scholes model, the Kou and Merton models achieve lower information criteria, indicating that their additional complexity is economically justified by genuinely improved explanatory power. The findings confirm that incorporating jumps into the price dynamics is essential for accurately modeling cryptocurrency options, and that this structural feature cannot be compensated by simpler diffusion-based frameworks.

While the error metrics presented above demonstrate differences in pricing performance across models, they do not formally assess whether these differences are statistically significant. To address this, we employ two complementary statistical tests: the Diebold--Mariano (DM) test for pairwise model comparisons and the Model Confidence Set (MCS) procedure for joint evaluation of multiple models  \citep{diebold2002comparing, hansen2011model}.

The DM test evaluates the null hypothesis that two models have equal predictive accuracy against the alternative that one model performs better. We compute the test statistic using squared pricing errors as the loss function and apply heteroskedasticity and autocorrelation consistent (HAC) standard errors to account for potential dependence in the error structure across strike prices. For BTC options, pooling all maturities, the results provide strong statistical evidence that jump-diffusion models significantly outperform simpler alternatives. The Kou model yields statistically significantly lower pricing errors than Black--Scholes, Merton, and Bates. Similarly, Merton significantly outperforms Black--Scholes. For ETH options, Merton significantly outperforms Black--Scholes and Bates.

The Model Confidence Set (MCS) procedure extends pairwise testing by identifying a set of models that are statistically indistinguishable as the best-performing models. The MCS eliminates inferior models sequentially based on an equivalence test, ultimately producing a superior set of models containing all models that cannot be rejected as suboptimal at a given significance level. We apply the MCS procedure with a 10\% significance level using the squared error loss function. For BTC options, the superior set consists of the Kou and Merton models, with all other models being statistically eliminated as inferior. For ETH options, the superior set includes Merton, Bates, and Kou, while Black--Scholes, Variance Gamma, and Heston are eliminated. These formal statistical tests provide robust confirmation of our main findings. 

The empirical findings presented in this paper not only demonstrate the statistical superiority of jump–diffusion and stochastic–volatility–with–jumps models, but also carry important economic implications regarding the behaviour of cryptocurrency markets. 

The consistently high jump intensities and large jump size parameters obtained for both BTC and ETH indicate that cryptocurrency prices are strongly affected by discontinuous movements rather than smooth diffusive dynamics. This finding aligns with previous studies documenting the prevalence of sharp price dislocations in digital asset markets, driven by liquidity shocks, exchange outages, forced liquidations in leveraged positions, and rapid reactions to news \citep{hou2020pricing, brini2024pricing}. Unlike traditional markets, where price jumps often reflect macroeconomic announcements or earnings reports, the cryptocurrency market is more tightly coupled with microstructure frictions, such as funding rate imbalances, liquidation cascades on derivatives exchanges, and sentiment-driven trading. These features naturally lead to heavier tails and sudden volatility bursts, which jump–diffusion models are designed to capture.

The parameter estimates obtained for Merton, Kou, Heston, and Bates models provide insight into market participants’ perceptions of tail risk and volatility dynamics. For both BTC and ETH, the high values of jump intensity across maturities suggest that market participants assign substantial probability to extreme price moves. The large jump size parameters observed in the Kou and Bates model calibrations further imply that these extreme movements are expected to be asymmetric, with downside jumps typically larger or more frequent than upside ones—a phenomenon consistent with the leverage effect and with empirical evidence on crash-risk asymmetry in crypto markets \citep{konczal2024tail}.

The stochastic volatility parameters also convey important information. The elevated volatility-of-volatility estimates and near-zero correlations between price and volatility suggest that volatility dynamics are themselves highly unstable and not strongly mean-reverting, in contrast to what is observed in equity or commodity markets. Similar findings have been reported in the literature, with Bitcoin exhibiting volatility levels four to five times higher than those of major equity indices \citep{nzokem2024bitcoin}. These calibrated characteristics highlight a market structure in which uncertainty remains persistently high and where volatility shocks propagate more rapidly than in traditional financial assets.

The results can also be viewed in the context of fundamental and behavioural factors shaping cryptocurrency markets. Prior work has shown that digital asset prices are influenced by blockchain fundamentals, network activity, miner revenues, and on-chain flows \citep{kristoufek2015main, jani2017overview}, but are equally driven by speculative cycles, sentiment, and liquidity conditions \citep{chen2020bitcoin}. Periods of elevated jump intensity frequently coincide with regulatory announcements, security breaches, liquidations of large leveraged positions, or significant updates to blockchain ecosystems. These events create sudden repricing pressures that models without jumps cannot capture. The steep term-structure differences observed in our calibrations also indicate that expectations of risk vary sharply over time, possibly reflecting uncertainty around upcoming protocol upgrades, macroeconomic conditions, or market-wide liquidity cycles.

\section{Conclusions}
\label{sec:conclusions}

In 2008, Satoshi Nakamoto introduced Bitcoin, which is a decentralized digital currency powered by blockchain technology \citep{nakamoto2008bitcoin}. Since then, Bitcoin has become the first and most widely recognized cryptocurrency. This success led to the development of thousands of other cryptocurrencies, including Ether. As these cryptocurrencies have gained popularity, they have also become significant assets for trading and investment. A growing interest in derivatives such as options on Bitcoin and Ether futures has emerged. These options are used to manage risk, but their unique characteristics require specific modeling and pricing.

Pricing options on cryptocurrency futures presents unique challenges due to the high volatility and sudden price jumps. In this study, we evaluated the performance of several models, namely: Black--Scholes, Variance Gamma, Merton Jump Diffusion, Kou, Heston and Bates. The analysis involved calibrating these models using out-of-the-money market data for vanilla call options on Bitcoin and Ether futures contracts and calculating pricing errors using various metrics such as RMSE, MAE, MAPE, and MSLE for eight different maturities. To capture a specific market scenario, we focused our analysis on a single trade date, March 11, 2024, a period of extreme turbulence in the cryptocurrency market. This allows us to assess how well these models perform under highly volatile conditions with significant price jumps. 

 An important limitation of this study is that our analysis focuses on a single trading day (March 11, 2024), chosen specifically to examine model performance under extreme market turbulence. While this provides rigorous testing under stress conditions, it limits the generalizability of our findings to other market regimes. Multi-date calibration and rolling-window analysis would provide a more comprehensive assessment of model stability across different market conditions.

The results show that the Black--Scholes model achieved the highest errors for all considered metrics. For BTC options, the Black–Scholes model yielded an RMSE of 1180, MAE of 854, and MAPE of 9.23\%. This confirms that the BS model is not suitable for pricing cryptocurrency options accurately. In comparison, models incorporating jump components and stochastic volatility, such as Kou and Merton Jump Diffusion, show significant improvements in pricing accuracy. Specifically, the Kou model achieved the lowest errors for BTC options with an RMSE of 649, MAE of 258, and MAPE of 2.64\%. This proves that the Kou model provides the most accurate pricing for BTC options among all the analyzed models.

Similarly, for ETH options, the Kou model also performed very well, with an RMSE of 30.8, MAE of 18.2, and MAPE of 4.12\%. However, the Stochastic Volatility with Jumps model performed even better, achieving the lowest errors across all metrics, including an impressive MAPE of just 1.9\%. The Merton Jump Diffusion model also provided good results, with an RMSE of 25.9, MAE of 16.6, and MAPE of 4.23\%. It confirms the relevance of including jump components in accurately pricing cryptocurrency derivatives.

The obtained results clearly highlight that models incorporating both jumps and stochastic volatility better capture the unique characteristics of cryptocurrency markets, which are marked by extreme price fluctuations and irregular behavior. While no single model can be considered a universal choice across all maturities and error metrics, the Kou and SVJ models stand out as the best-performing models for both BTC and ETH options. These models significantly improve the accuracy of option pricing compared to the traditional Black–Scholes model, making them more suitable for pricing options on highly volatile and dynamic cryptocurrency markets.

In addition to the empirical findings presented in this study, our results open several promising directions for future research. First, since cryptocurrency markets exhibit pronounced volatility clustering and highly irregular short-term movements, models based on rough volatility or more flexible stochastic volatility specifications may offer additional improvements over the classical Heston and Bates models. Second, machine learning models could help capture nonlinear effects and microstructural patterns that are difficult to model analytically, particularly during periods of extreme turbulence. Another natural extension involves the use of high-frequency data and order book information, which may allow for more accurate modelling of short-term jumps, liquidity frictions, and the shape of the implied volatility surface. Future research could also examine the hedging performance of these models by analyzing Greeks and conducting dynamic hedging simulations using high-frequency time-series data, which would provide insights into risk management applications beyond cross-sectional pricing accuracy.

The findings presented here also have several practical implications for traders and risk managers. Since models incorporating jumps and stochastic volatility significantly outperform Black--Scholes for both BTC and ETH options, relying solely on diffusion-based frameworks may lead to persistent mispricing and an underestimation of tail risk. For market participants engaged in hedging or managing leveraged positions, jump–diffusion and stochastic volatility with jumps models provide hedge ratios and risk measures that better reflect the true behaviour of cryptocurrency markets. In particular, during episodes of sharp price dislocations, such as those observed on 11 March 2024, models capable of capturing sudden jumps can support more robust hedging strategies.

\section*{Acknowledgement}

The author gratefully acknowledges the valuable remarks provided by Krzysztof Burnecki and Michał Balcerek.
The author also thanks the Anonymous Reviewer for insightful comments that significantly improved the paper. The work was supported by NCN Grant No. 2022/47/B/HS4/02139.

\bibliographystyle{ws-ijtaf}
\bibliography{bibliography}
\end{document}